\documentclass[aip, amsmath, amssymb,reprint]{revtex4-2}
\usepackage{graphicx}
\usepackage{dcolumn}
\usepackage{gensymb}
\usepackage{xcolor}
\usepackage{hyperref}
\usepackage{float}
\usepackage{outlines}
\hypersetup{colorlinks=true,pdfborder={0 0 0},}
\usepackage{enumitem}
\usepackage{upgreek}
\usepackage[normalem]{ulem}
\usepackage{bm}
\usepackage{color,soul} 

\usepackage{xr}
\makeatletter
\newcommand*{\addFileDependency}[1]{
  \typeout{(#1)}
  \@addtofilelist{#1}
  \IfFileExists{#1}{}{\typeout{No file #1.}}
}
\makeatother

\newcommand*{\myexternaldocument}[1]{
    \externaldocument{#1}
    \addFileDependency{#1.tex}
    \addFileDependency{#1.aux}
}
\usepackage{xr}
\makeatletter

\myexternaldocument{SI}

\begin{document}
\title{Segment-chirped periodically poled lithium niobate waveguides for broadband supercontinuum generation}

\author{Yue Li} 
\affiliation{Department of Materials Science and Engineering, National University of Singapore, Singapore 117575, Singapore}
\affiliation{College of Electronics and Information Engineering, Sichuan University, Chengdu 610065, China}

\author{Xiaodong Shi}
\affiliation{A$^\ast$STAR Quantum Innovation Centre (Q.InC), Agency for Science, Technology and Research (A$^\ast$STAR), Singapore 138634, Singapore}

\author{Sakthi Sanjeev Mohanraj}
\affiliation{Department of Materials Science and Engineering, National University of Singapore, Singapore 117575, Singapore}

\author{Mengyao Zhao}
\affiliation{Department of Materials Science and Engineering, National University of Singapore, Singapore 117575, Singapore}

\author{Xu Chen}
\affiliation{Department of Materials Science and Engineering, National University of Singapore, Singapore 117575, Singapore}

\author{Xuan Mao}
\affiliation{School of Electrical and Electronic Engineering, Nanyang Technological University, Singapore 639798, Singapore}

\author{Qijie Wang}
\affiliation{School of Electrical and Electronic Engineering, Nanyang Technological University, Singapore 639798, Singapore}

\author{Shouhuan Zhou}
\affiliation{College of Electronics and Information Engineering, Sichuan University, Chengdu 610065, China}

\author{Guoliang Deng}
\affiliation{College of Electronics and Information Engineering, Sichuan University, Chengdu 610065, China}

\author{Di Zhu}
\email{dizhu@nus.edu.sg}
\affiliation{Department of Materials Science and Engineering, National University of Singapore, Singapore 117575, Singapore}
\affiliation{A$^\ast$STAR Quantum Innovation Centre (Q.InC), Agency for Science, Technology and Research (A$^\ast$STAR), Singapore 138634, Singapore}
\affiliation{Centre for Quantum Technologies, National University of Singapore, Singapore 117543, Singapore}

\begin{abstract}
Supercontinuum generation is a key technology in nonlinear optics, supporting a wide range of applications in frequency metrology and spectroscopy. Integrated photonics offers a promising route toward compact and efficient supercontinuum sources, yet extending the bandwidth while maintaining high spectral flatness remains a central challenge. Here we demonstrate an integrated broadband supercontinuum source based on segment-chirped periodically poled lithium niobate (SC-PPLN) nanophotonic waveguides. By discretizing the chirped poling profile into independently optimized segments, this approach enables high-fidelity ferroelectric domain inversion with near-ideal duty cycles and establishes broadband quasi-phase matching, overcoming the domain inhomogeneity and efficiency limitations commonly encountered in conventional chirped poling. The engineered phase-matching landscape supports efficient wavelength conversion and simultaneous activation of multiple second- and third-order nonlinear processes. Experimentally, we achieve a spectrally flat supercontinuum spanning three optical octaves, from 320 nm in the ultraviolet to 2600 nm in the mid-infrared. These results establish segment-chirped poling as a practical strategy for broadband wavelength conversion and supercontinuum generation in integrated photonics.
\end{abstract}

\maketitle

\section{Introduction}

Broadband wavelength conversion and supercontinuum generation play crucial roles in modern nonlinear optics, underpinning applications ranging from frequency metrology and spectroscopy to optical coherence tomography and ultra-fast signal processing \cite{drexler2004ultrahigh,hartl2001ultrahigh,nishizawa2004real,sotobayashi2001bi,sotobayashi2002wavelength, hu2018single}. 
In particular, coherent light sources spanning from the ultraviolet (UV) to the mid-infrared (mid-IR) are critically important for molecular fingerprinting, nonlinear microscopy, environmental sensing, and on-chip frequency synthesis \cite{diddams2007molecular,timmers2018molecular,min2011coherent,du2019mid,gaeta2019photonic}. 
Despite substantial progress, achieving such extreme spectral coverage with high efficiency, spectral flatness, and chip-scale integration remains a longstanding challenge.

Supercontinuum has traditionally relied on specialized optical fibers, where long interaction length and engineered dispersion enable octave-spanning spectra \cite{dudley2006supercontinuum,birks2000supercontinuum,hu2010maximizing}. 
However, these approaches often suffer from limited nonlinear efficiency and large footprints. 
Integrated photonic platforms offer a compelling alternative, providing strong optical confinement, flexible dispersion engineering, and compatibility with wafer-scale fabrication. 
Most prior demonstrations of broadband coherent light sources in integrated platforms, such as silica, silicon nitride, aluminum nitride, and silicon carbide, primarily rely on third-order nonlinearity ($\chi^{(3)}$), exploiting soliton dynamics and dispersive-wave emission in dispersion-engineered waveguides \cite{johnson2015octave,yan2026simplified,psaila2007supercontinuum,li2024efficient,afridi20244h}.
However, $\chi^{(3)}$-based approaches can only access a restricted set of nonlinear processes, thereby limiting spectral expansion and constraining the achievable bandwidth.

In contrast, thin-film lithium niobate (TFLN) integrated platform supports both strong $\chi^{(2)}$ and $\chi^{(3)}$ interactions within a wide window of transparency, enabling coherent interplay between multiple nonlinear processes and offering a pathway to spectral coverage beyond what is achievable with $\chi^{(3)}$ interactions alone \cite{zhu2021integrated,peng2025three,hamrouni2024picojoule,li20262,zhou2025quadratic,jankowski2020ultrabroadband,yu2019coherent,fan2025spectral,lu2019octave,wu2024visible,Tang:25,10.1063/5.0028776,ayhan2025fabrication,ludwig2025mid,tang2025chip,fang2026broadband1,gao2025tightly}.
Realizing this potential, however, critically depends on achieving efficient and simultaneous $\chi^{(2)}$ phase matching across a broad spectral range. 
Conventional periodic poling enables flexible quasi-phase matching (QPM), but only supports discrete wavelengths with narrow bandwidth \cite{shi2025integrated,bollmers2025segmented}.
While chirped or aperiodic poling schemes can extend the phase-matching bandwidth, due to practical fabrication nonidealities, they often suffer from imperfect domain inversion, duty-cycle fluctuations, and reduced effective nonlinear overlap. 
These issues collectively degrade conversion efficiency and spectral uniformity, posing a major obstacle to realizing efficient, broadband, and spectrally flat $\chi^{(2)}$-assisted wavelength conversion and supercontinuum generation on chip.

\begin{figure*}[htbp]
\centering 
\includegraphics[width = 6.3in]{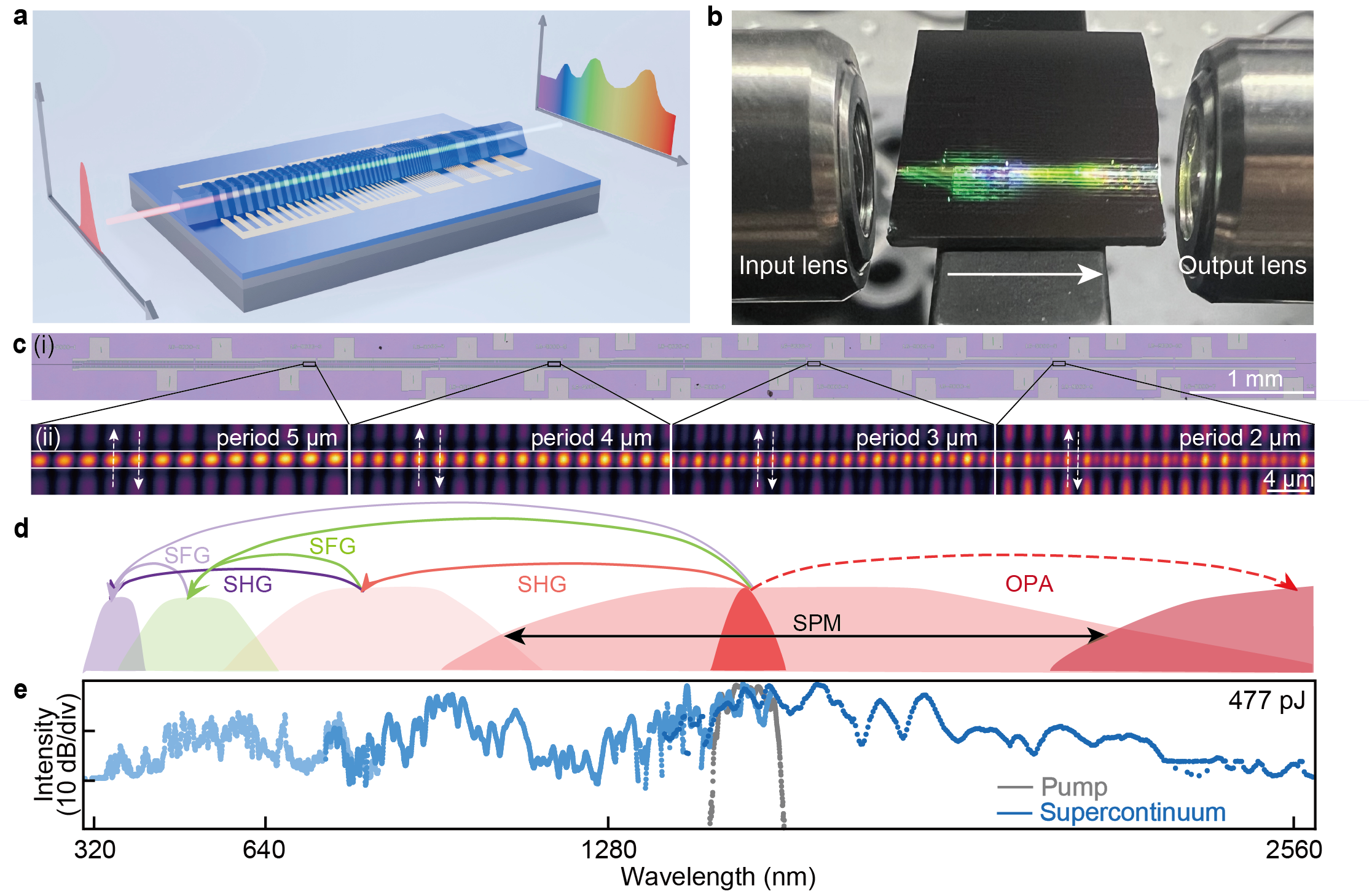}
\caption{\textbf{Segment-chirped periodically poled lithium niobate (SC-PPLN) nanophotonic waveguide for broadband and coherent wavelength conversion.} \textbf{a,} Schematic of segment-chirped periodically poled lithium niobate nanophotonic waveguide for supercontinuum generation. \textbf{b,} Photograph of the device producing supercontinuum light. \textbf{c,} (i) Optical micrograph of the segment-chirped periodically poled waveguide, and (ii) second-harmonic microscopic imaging of the poling regions. \textbf{d,} Schematic of the spectral broadening mechanisms for supercontinuum generation, including second-harmonic generation (SHG), sum-frequency generation (SFG), optical parametric amplification (OPA), and self-phase modulation (SPM). \textbf{e,} Measured supercontinuum spectrum (blue line) from UV to mid-infrared, corresponding to three optical octaves. The gray line is pump pulse spectrum.}
\label{fig1}
\end{figure*}

Here, we experimentally demonstrate an integrated broadband supercontinuum source in a TFLN nanophotonic waveguide enabled by segment-chirped periodic poling (Fig.~\ref{fig1}a–c). 
In this approach, the chirped periodic poling profile is discretized into multiple segments with distinct poling periods, allowing independent optimization of the poling conditions, including pulse amplitude, duration, and number of pulses, for each segment. 
This strategy enables full-depth ferroelectric domain inversion with a near-ideal 50\% duty cycle across the entire poled waveguide, preserving a high effective $\chi^{(2)}$ nonlinearity over a broad spectral range. 
As a result, multiple $\chi^{(2)}$-mediated processes, including second-harmonic generation (SHG), sum-frequency generation (SFG), optical parametric amplification (OPA), and cascaded $\chi^{(2)}$ interactions, are efficiently excited alongside $\chi^{(3)}$ nonlinear effects. 
The coexistence of these efficient nonlinear processes gives rise to a synergistic spectral broadening mechanism fundamentally distinct from Kerr-only supercontinuum generation (Fig.~\ref{fig1}d).

Using this technique, we achieve efficient generation of second-, third-, and fourth-harmonic light, together with a broadband spectrum spanning from 320~nm in UV to 2600~nm in mid-IR, corresponding to three optical octaves with a flat power spectral density (Fig.~\ref{fig1}e). 
The measured broadband wavelength-conversion efficiency closely approaches theoretical predictions, highlighting the high quality and uniformity of the segment-chirped periodic poling. These results establish segment-chirped periodically poled lithium niobate (SC-PPLN) nanophotonic waveguides as a viable and versatile platform for integrated frequency combs, broadband spectroscopy, and advanced nonlinear signal processing on a single photonic chip.

\section{Results}

\paragraph{Device principle.}

\begin{figure*}[htbp]
\centering 
\includegraphics[width = 6.3in]{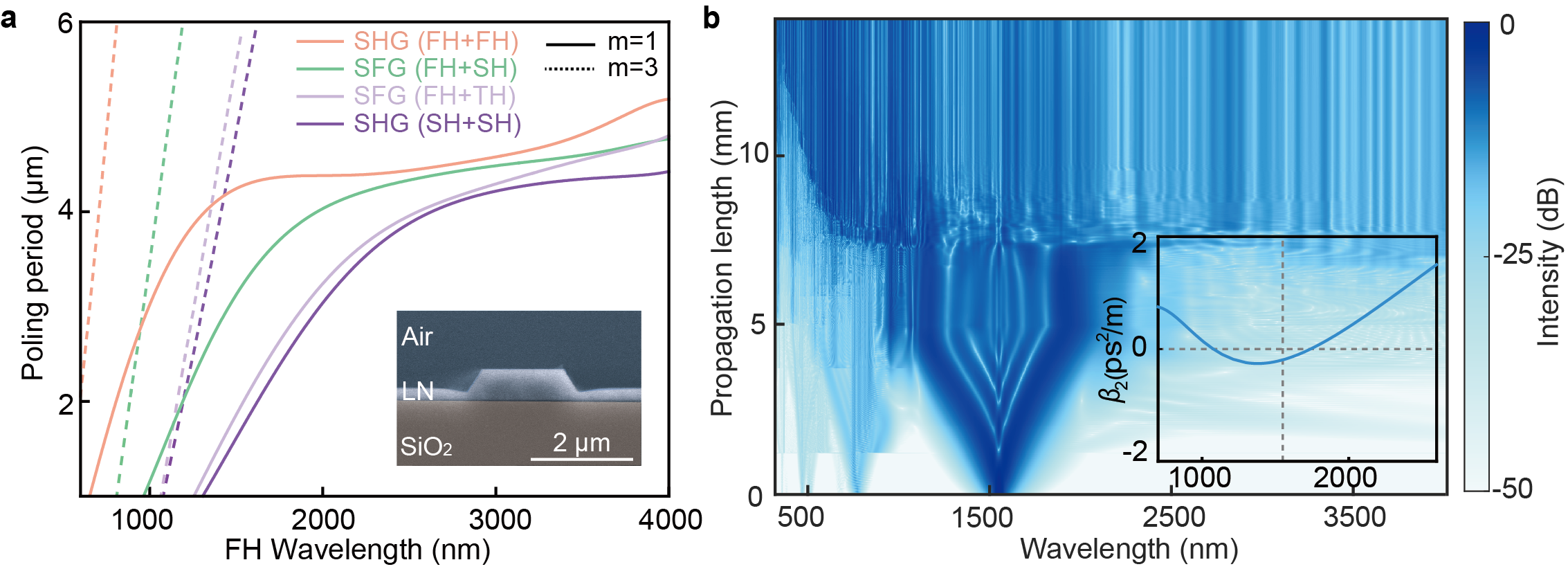}
\caption
{\textbf{Simulation of wavelength conversion and supercontinuum generation.} \textbf{a,} Quasi-phase-matched poling period as a function of fundamental-harmonic (FH) wavelength for various $\chi^{(2)}$ nonlinear processes. Solid lines show first-order phase matching, while dashed lines show third-order phase matching. SH: second-harmonic, TH: third-harmonic. Inset: false-color cross-sectional scanning electron micrograph of the fabricated waveguide. \textbf{b,} Simulated spectral evolution in a 14 mm long segment-chirped poled TFLN waveguide, consisting of a 5 mm unpoled section and a 9 mm segment-chirped periodically poled section, with a pump pulse energy of 477 pJ. The inset shows simulated group velocity dispersion indicating that the pump at 1560 nm is in the near-zero anomalous dispersion regime.} 
\label{fig2}
\end{figure*}

The key design strategy is to enable broadband phase matching for  $\chi^{(3)}$ and $\chi^{(2)}$ nonlinear interactions through careful dispersion engineering and high-quality chirped poling, respectively. The device is designed on 600~nm-thick $x$-cut MgO-doped TFLN. 
The waveguide has a width of 1600~nm and an etch depth of 350~nm (inset of Fig.~\ref{fig2}a). 
Numerical simulations indicate near-zero anomalous group-velocity dispersion around 1560 nm (inset of Fig.~\ref{fig2}b), which helps to balance self-phase modulation (SPM) and avoid excessive temporal pulse broadening that would otherwise suppress nonlinear efficiency.
The soliton number is designed to lie in the range of $1 \le N \le 10$, ensuring that nonlinear effects dominate over dispersion during propagation and enabling efficient pulse self-compression \cite{boyd2008nonlinear}.
The length of the unpoled TFLN waveguide section is set to 5 mm, that effectively allows $\chi^{(3)}$ based SPM to pre-broaden the pump spectrum, and seed subsequent $\chi^{(2)}$ interactions.

We then simulate QPM conditions for SHG, cascaded SHG, and SFG, involving the fundamental and high-order harmonic fields in the waveguide (Fig. \ref{fig2}a, see simulation details in Methods).
The required poling period varies from 1 \textmu m to 5 \textmu m for the first-order QPM processes at fundamental wavelengths between 600 nm and 4000 nm. 
In the short wavelength range, the required QPM periods decrease rapidly owing to strong material and waveguide dispersion, approaching sub-micrometer scale.
Direct realization of sub-micrometer poling is challenging due to lateral domain broadening and duty-cycle distortion during electrical poling. 
To mitigate this problem, third-order QPM schemes are partially employed, thereby providing efficiency compensation in the short wavelength range (see analysis in Supplementary S1 and Fig.~\ref{figs1}) \cite{fejer1992quasi}.

The total device length is 14 mm, consisting of a 5 mm unpoled section and a subsequent 9 mm SC-PPLN section with poling period spanning from 6 \textmu m to 1\textmu m.
When fabricating the device, we find that it is difficult to find a universal poling condition to achieve high-quality poling of waveguides with varying periods.
For example, the large-period region requires high voltage to overcome broader energy barriers, but the small-period region will be much over-poled under the same poling condition, due to highly concentrated electric fields and fast inversion response.
Thus, the chirped poling region is discretized into 10 segments, by discontinuing the metal combs for electrical poling, within which the variation of the poling period is restricted to be 0.5 \textmu m (Fig. \ref{fig1}c(i)). 
This segmented approach allows effective control of electrical poling of varying poling periods, and enables high-quality poling in terms of a uniform duty cycle of 0.5 and full-depth domain inversion. 
As confirmed by SH microscopic images (Fig.~\ref{fig1}c(ii)), uniform and well-defined domain inversion with $\sim50\%$ duty cycle is achieved along the whole waveguide.  

We show that the SC-PPLN waveguide enables broadband $\chi^{(2)}$ phase matching through simulation (see Supplementary Fig. \ref{figs3}).
Simulations of spectral broadening in the designed structure reveal that the unpoled section is dominated by $\chi^{(3)}$-driven SPM, while the segment-chirped periodically poled region activates multiple $\chi^{(2)}$ nonlinear interactions, including SHG, SFG, OPA, as well as their cascaded processes (Fig.~\ref{fig2}b).
Together, the dispersion-engineered waveguide and the segment-chirped QPM structure establish the physical basis for efficient and broadband wavelength conversion and flat coherent spectral expansion from UV to mid-IR, with pump pulse energy above 200 pJ (see Supplementary Fig. \ref{figs2}).

\paragraph{Broadband and efficient wavelength conversion.}

\begin{figure*}[htbp]
\centering 
\includegraphics[width = 6.3in]{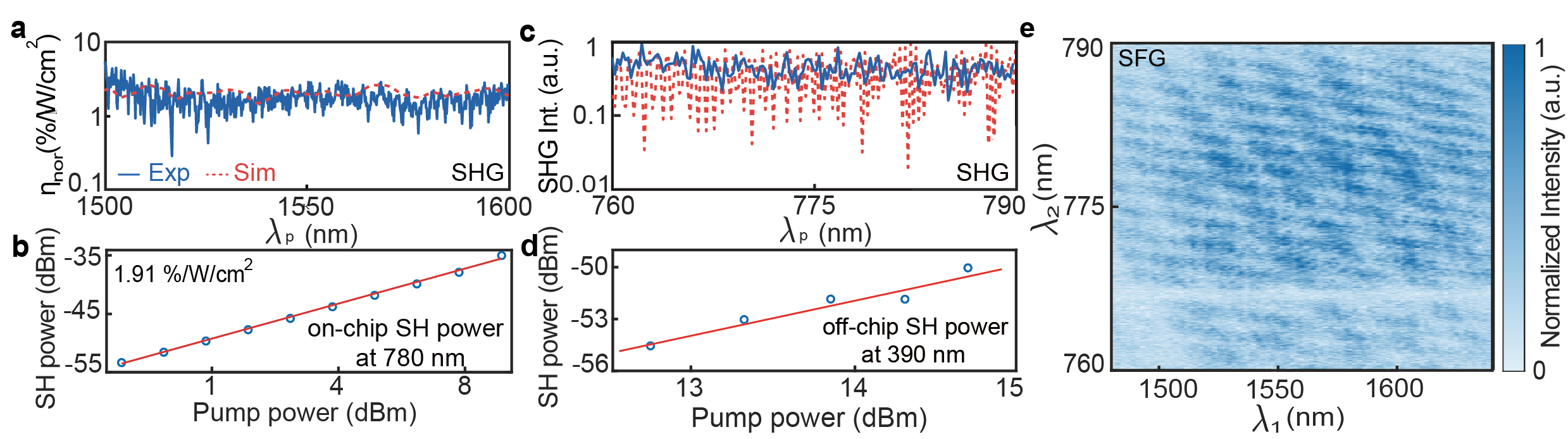}
\caption{\textbf{Broadband and efficient wavelength conversion.}
\textbf{a,} Measured (blue solid line) and simulated (red dashed line) results of the normalized SHG efficiency based on the 9 mm-long chirped poling region from 1500 nm to 1600 nm. \textbf{b,} Measured on-chip SHG power (blue circles) as a function of pump power at 1560 nm, with quadratic fitting (red line) yields normalized conversion efficiency of 1.9~\%~W$^{-1}$cm$^{-2}$. \textbf{c,} Measured (blue solid line) and simulated (red dashed line) results of the normalized SHG intensity from 760 nm to 790 nm. \textbf{d,} Measured off-chip SHG power (blue circles) as a function of pump power at 780 nm with quadratic fitting (red line). \textbf{e,} Measured SFG intensity mapping between 1480 nm - 1640 nm and 760 nm - 790 nm.}
\label{fig3}
\end{figure*}

We first characterize the wavelength-conversion capability enabled by the segment-chirped $\chi^{(2)}$ phase matching. 
Continuous-wave (CW) pumps are employed to probe the SHG and SFG processes.
Using a telecom-band tunable laser as the pump, broadband SHG is observed (Fig.~\ref{fig3}a). 
The experimentally measured wavelength-dependent SHG efficiency shows good agreement with the theoretical simulations, verifying that the device enables broadband and efficient $\chi^{(2)}$ frequency conversion (see analysis in Supplementary S2). 
At a pump wavelength of 1560 nm, a normalized on-chip conversion efficiency of 1.9~\%~W$^{-1}$~cm$^{-2}$, approaching the theoretical value (2.1~\%~W$^{-1}$~cm$^{-2}$), is extracted from the quadratic power dependence measurement (Fig.~\ref{fig3}b), indicating the SHG nature. 

The SHG response is further characterized in the visible regime by pumping the waveguide with wavelengths ranging from 760 to 790 nm. 
Owing to the difficulty of accurately calibrating waveguide coupling losses in UV, the absolute on-chip SHG conversion efficiency cannot be reliably extracted in this spectral range. 
Instead, the experimentally measured SHG spectrum is qualitatively compared with numerical simulations (Fig.~\ref{fig3}c). 
A clear quadratic dependence of the generated signal on the pump power is observed in the measurement (Fig.~\ref{fig3}d), confirming the second-harmonic nature of the process. 
Notably, despite the increased fabrication sensitivity associated with short and widely varying poling periods, SHG persists across the visible and near-UV range, underscoring the robustness of the segmented poling strategy for short-period QPM.

In addition to SHG responses, SFG between telecom and visible wavelengths is characterized by synchronously sweeping two tunable CW lasers, one around 1550 nm and the other around 775 nm (Fig.~\ref{fig3}e).
The characteristic anti-diagonal spectral feature, as well as the SFG signal generation in the whole measured spectral range, well matches simulated results (see Supplementary Fig. \ref{figs4}) and provides direct evidence of simultaneous phase matching across a wide spectral range.
These results jointly demonstrate that the segment-chirped poling strategy enables efficient and continuous wavelength conversion spanning at least two optical octaves, enabling broadband frequency mixing and coherent spectral synthesis on chip.

\paragraph{Three-octave supercontinuum generation.}

\begin{figure*}[htbp]
\centering 
\includegraphics[width = 6.3in]{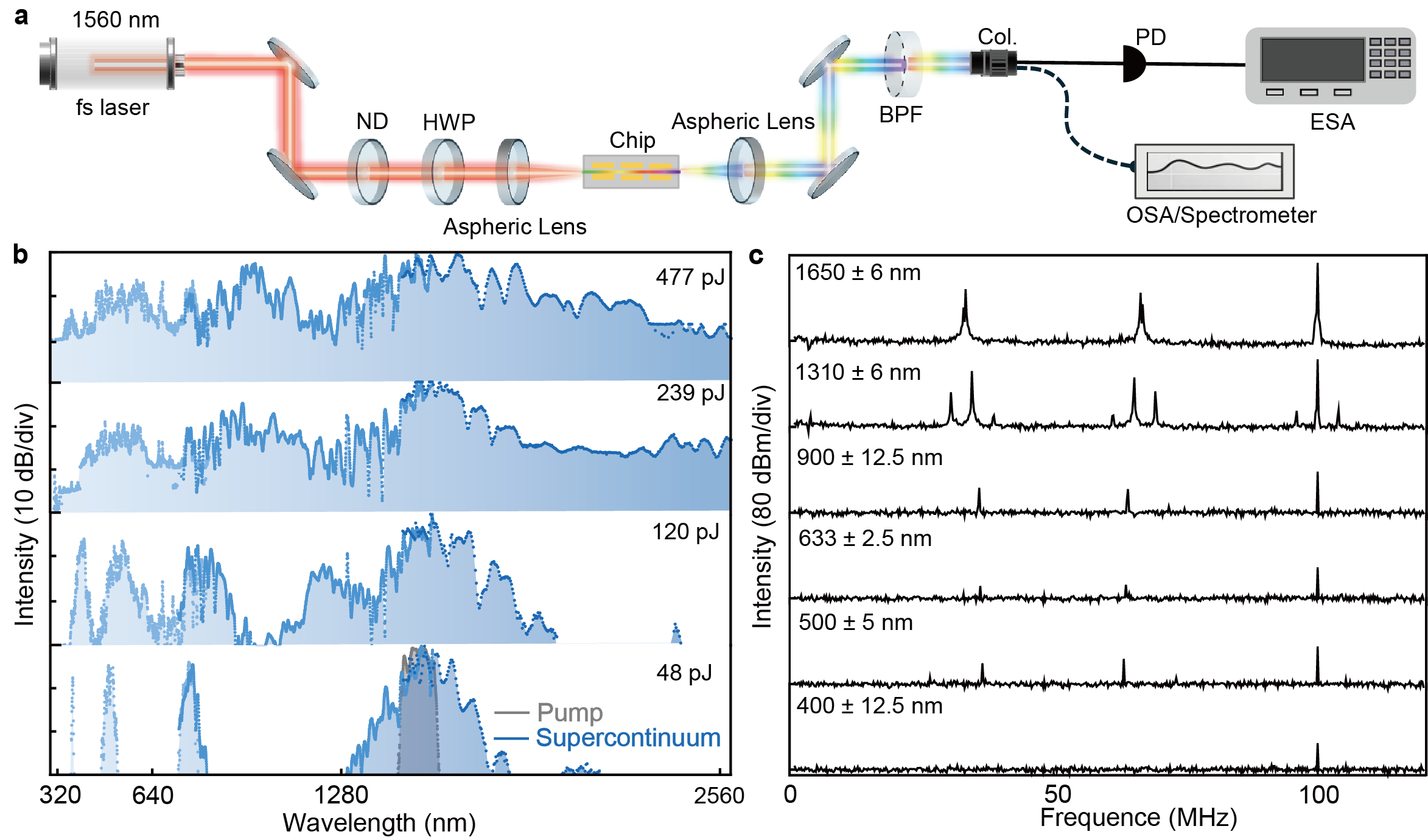}
\caption{\textbf{Three-octave supercontinuum generation from ultraviolet to mid-infrared.} \textbf{a,} Measurement setup schematic. fs laser: femtosecond laser; ND: neutral density filter; HWP: half-wave plate; BPF: bandpass filter; Col.:fiber collimator; PD: photodetector; ESA: electrical spectrum analyzer; OSA: optical spectrum analyzer. \textbf{b,} Supercontinuum spectra with increasing pump pulse energy (from bottom to top: 48 pJ, 120 pJ, 239 pJ and 477 pJ), offset by 30 dB. \textbf{c,} The RF beatnotes with pump pulse energy of 477 pJ at various wavelengths (400 nm, 500 nm, 633 nm, 900 nm, 1310 nm, and 1650 nm, indicating coherent supercontinuum generation.}
\label{fig4}
\end{figure*}

Building on the demonstrated broadband $\chi^{(2)}$ response, we next investigate the nonlinear dynamics under femtosecond pulse pumping (Fig.~\ref{fig4}a). 
An 80 fs pulsed laser with a center wavelength at 1560 nm is coupled into the waveguide, with on-chip pulse energy increasing from 48 pJ to 477 pJ. 
The output spectra are shown in Fig.~\ref{fig4}b.
At low pulse energies, distinct spectral features are observed, corresponding to SHG, SFG, and cascaded SHG processes. 
As the pulse energy increases, $\chi^{(3)}$-mediated SPM in the unpoled section broadens the fundamental spectrum, which increasingly overlaps with $\chi^{(2)}$-generated harmonics, SFG, and OPA products in the poled region. 
Eventually, discrete spectral features merge into a continuous and spectrally flat supercontinuum with intensity fluctuation within 20 dB, spanning from 320~nm in UV to 2600~nm in mid-IR, corresponding to three optical octaves (Fig.~\ref{fig1}e).
Numerical simulations incorporating both $\chi^{(2)}$ and $\chi^{(3)}$ nonlinearities reproduce the observed spectral evolution (Fig.~\ref{fig2}b). 
Importantly, the simulations predict the emergence of mid-infrared wavelength components up to 4 \textmu m, suggesting that the measured long-wavelength cutoff is imposed by instrument limits rather than phase-matching constraints.

We also measure the radio-frequency (RF) beatnotes at different wavelengths and pump pulse energies.  
Figure \ref{fig4}c shows the results at 477 pJ, and those at other pump pulse energies can be found in Supplementary Fig. \ref{figs5}. 
Narrow and stable beatnotes at 100 MHz are observed across visible and near-infrared spectral slices, confirming that the generated supercontinuum is coherent.
At relatively low pump pulse energies (48~pJ and 120~pJ), no pronounced RF beatnote signals are detected in most spectral bands (except for the original pump repetition rate of 100 MHz), indicating that the spectral components remain largely isolated and different comb families do not significantly overlap. 
When the pump pulse energy is increased above 200~pJ, additional RF beatnote signals appear near the carrier-envelope offset frequency and its complementary folding components from visible to infrared, suggesting that spectral overlap between different comb families occurs within these regions, thereby leading to the formation of a continuous and gap-free supercontinuum (see analysis in Supplementary S3).

Together, these results demonstrate that the SC-PPLN offers a practical and reliable approach for implementing broadband supercontinuum generation on chip, in which $\chi^{(2)}$ phase matching and $\chi^{(3)}$ nonlinearities act synergistically to produce coherent light spanning from UV to mid-IR on an integrated photonic chip.

\section{Discussion}

This work establishes segment-chirped $\chi^{(2)}$ phase matching as an effective and scalable strategy for broadband wavelength conversion and supercontinuum generation in TFLN nanophotonics. 
Unlike conventional continuously chirped periodic poling, which requires a single set of electrical poling conditions to accommodate a wide range of poling periods, the SC-PPLN discretizes the chirped profile into multiple independently optimized regions. 
This decouples the poling fidelity from the overall bandwidth requirement and enables full-depth ferroelectric domain inversion with a near-ideal 50\% duty cycle across the entire device, even in the presence of large period variations.

By preserving a high effective nonlinear coefficient over all segments, the segment-chirped architecture enables the efficient and simultaneous activation of multiple $\chi^{(2)}$-mediated processes spanning from UV to mid-IR. 
When combined with $\chi^{(3)}$ nonlinear effects supported by the dispersion-engineered waveguide, this dense phase-matching landscape gives rise to a synergistic nonlinear dynamics that differs fundamentally from Kerr-only supercontinuum generation. 
In particular, broadband SHG, SFG, OPA, and cascaded nonlinear processes collectively redistribute energy across widely separated spectral regions, resulting in both broad bandwidth and enhanced spectral flatness. 
In the present device, this manifests as a continuous spectrum with a 20 dB bandwidth exceeding three optical octaves. 
A comparison with previously reported results in thin-film lithium niobate and other integrated material platforms (Fig. \ref{fig5}, see detailed information in Supplementary Table \ref{tab}), highlights the demonstrated supercontinuum source in this work has broad bandwidth and high spectral flatness, enabled by segment-chirped periodic poling \cite{leo2014dispersive,singh2018octave,yoon2017coherent,grassani2019mid,porcel2017two,deniel2023visible,yan2026simplified,kuyken2020octave,fan2024supercontinua,dave2015dispersive,jung2021tantala,singh2020nonlinear,hammani2018octave,hwang2021supercontinuum,choi2016nonlinear,wang2025three,hamrouni2024picojoule,li20262,zhou2025quadratic,jankowski2020ultrabroadband,yu2019coherent,fan2025spectral,lu2019octave,wu2024visible,Tang:25,10.1063/5.0028776,ayhan2025fabrication,ludwig2025mid,tang2025chip,gao2025tightly,peng2025three}. 

\begin{figure*}[htbp]
\centering 
\includegraphics[width = 6.5in]{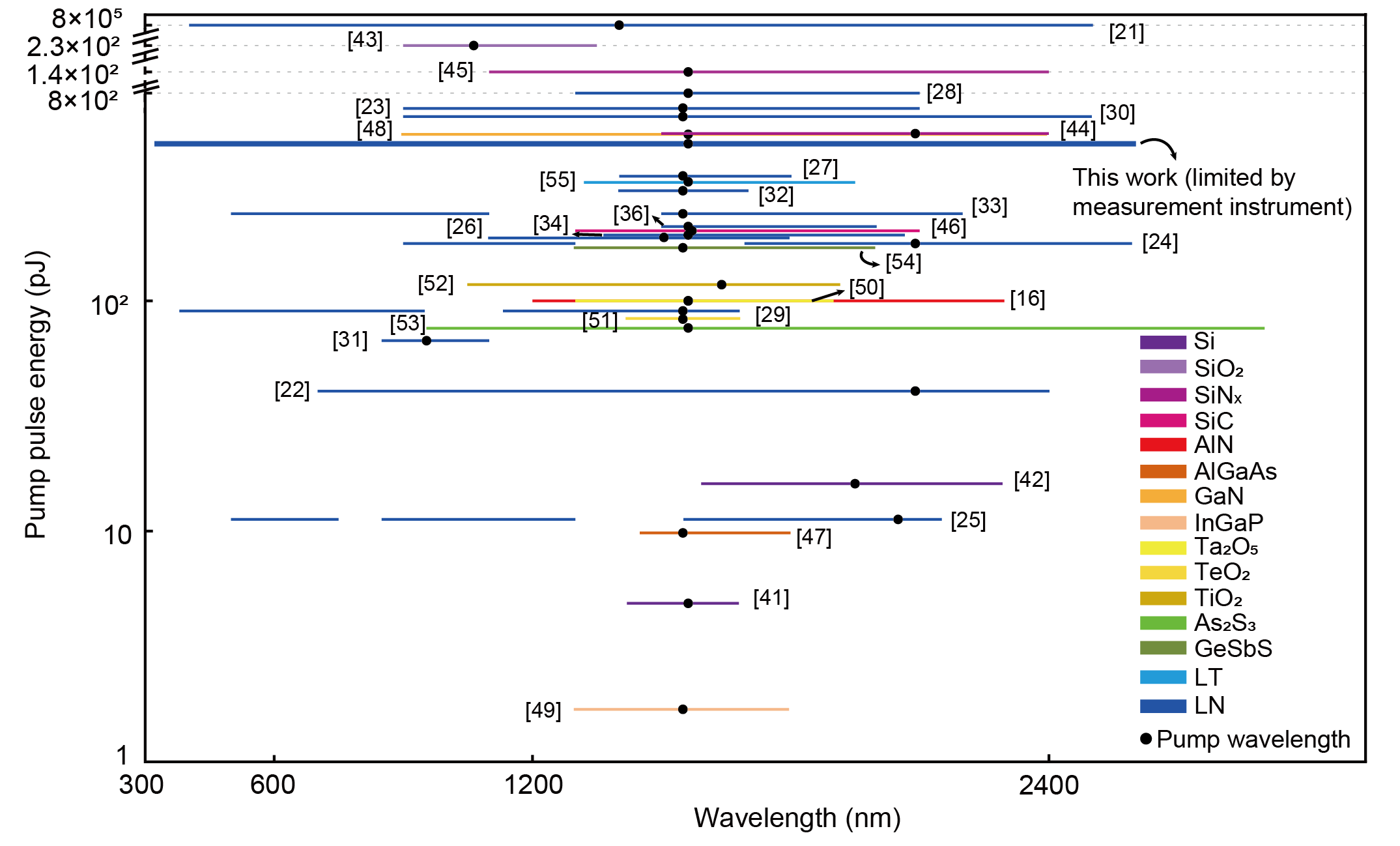}
\caption{\textbf{Comparison of supercontinuum generation with 20 dB bandwidth in various integrated photonic platforms.}}
\label{fig5}
\end{figure*}

Despite these advantages, several limitations of the current implementation merit discussion. 
Electrical poling in TFLN remains experimentally demanding, particularly for sub-micrometer poling periods required at short wavelengths, where fabrication tolerances become increasingly stringent. 
Although the segmented approach significantly improves domain fidelity and reproducibility, further advances in electrode design, pulse shaping, and localized or laser-assisted poling techniques could further enhance scalability and fabrication yield \cite{franken2025milliwatt,shi2024efficient}. 
Meanwhile, the generation of octave-spanning supercontinuum still requires a relatively high pump pulse energy in the current device.
Optimizing the segmented poling period distribution and the temporal evolution of cascaded nonlinear processes would reduce the pump threshold and improve conversion efficiency.
In addition, the long-wavelength extent of the measured supercontinuum is presently limited by the available detection bandwidth rather than by intrinsic phase-matching constraints. 
Future experiments employing mid-IR spectrometers, improved coupling schemes, and low-loss packaging will be essential to fully quantify the achievable spectral reach \cite{liang2022efficient}. 
Finally, while good coherence is observed across the generated spectrum, a more detailed characterization of phase noise and relative coherence between overlapping parametric sub-combs will be important for applications requiring absolute frequency stability, as well as for precision metrology \cite{helbing2002carrier}.

Looking forward, the demonstrated platform enables a broad range of applications that benefit from compact, coherent, and broadband light sources. 
On-chip generation of coherent radiation spanning from UV to mid-IR is particularly attractive for dual-comb spectroscopy, broadband molecular sensing, and integrated frequency metrology. 
The coexistence of harmonic generation, sum-frequency mixing, and parametric processes further enables direct wavelength conversion between otherwise inaccessible spectral regions, offering a compact and versatile interface between disparate photonic systems. 
In quantum photonics, the $\chi^{(2)}$ phase-matching landscape provided by segment-chirped periodic poling could be exploited for broadband squeezed light generation, entangled-photon generation, frequency-multiplexed quantum frequency conversion, and hybrid quantum-classical photonic architectures \cite{shi2025squeezed,fang2026broadband1}. 
More broadly, the SC-PPLN approach provides a general framework for engineering nonlinear interactions in $\chi^{(2)}$ nanophotonic platforms, opening new opportunities for fully integrated, multifunctional nonlinear photonic circuits.

\section*{Methods}
\paragraph{Device design.}
Quasi-phase matching is used to compensate for the phase mismatch among the waves propagating in nonlinear waveguides.
For a general three-wave mixing process (such as SFG), the $m$th-order QPM period is calculated by $\Lambda_m = \frac{2\pi m}{\Delta k}=\frac{2\pi m}{|k_{1}-k_{2}-k_{3}|}$,
where $k = \frac{2\pi n_{eff}}{\lambda}$ is the propagation constant (for SHG, $k_2 = k_3$), and $m = 1, 3$ indicates the first- and third-order of QPM, respectively. 

\paragraph{Device fabrication.}

The device is fabricated on a 600 nm thick MgO-doped $x$-cut TFLN chip (NanoLN). 
Prior to waveguide patterning, a 15 nm thick HfO$_2$ layer is deposited by atomic layer deposition (ALD), serving as an insulating buffer layer for subsequent electrical poling. 
For chirped poling, segmented comb-shaped electrodes are patterned on the HfO$_2$ surface using electron-beam lithography, followed by electron-beam evaporation of a 100 nm thick Ni layer and lift-off process. 
Electrical poling is performed with a sequence of high voltage pulses to reverse the ferroelectric domains in TFLN. 
The number of pulses and the pulse voltage are individually optimized for each poling section to ensure uniform domain inversion.

After poling, the waveguides are patterned using electron-beam lithography with MaN resist as the etch mask. 
The TFLN is then etched by inductively coupled plasma reactive-ion etching (ICP-RIE) using Ar$^+$ plasma. 
A top-view optical microscopic image and a cross-section scanning electron micrograph are shown in Fig. \ref{fig1}c(i) and inset of Fig. \ref{fig2}a, respectively.
SHG microscopic images of various chirped-poling sections are shown in Fig. \ref{fig1}c(ii).

\paragraph{Optical measurement setup for continuous-wave pumped wavelength-conversion measurements.}

For telecom and visible wavelength pumped SHG measurements (Fig. \ref{fig3}a-d), tunable CW lasers (Santec TSL-570, Toptica CTL-780) are used as the pump individually. 
The input polarization is adjusted using a fiber polarization controller and aligned to the transverse-electric (TE) polarization.
Output light is coupled into and out of the chip using a pair of lensed fibers. 
The fiber-to-chip coupling losses per facet are calibrated to be 4.4 dB at 1560 nm and 6.3 dB at 780 nm. 
The SHG signals are selected with short-pass filters and measured using germanium and silicon photodetectors for testing signals around 780 nm and 390 nm, respectively. 
For SFG measurements (Fig. \ref{fig3}e), light from two tunable laser sources are combined using a wavelength division multiplexer, and then is coupled into the device.
The SFG signal around 520 nm is detected after a short-pass filter with a silicon photodetector.
For wavelength-dependent SHG and SFG characterizations, the pump wavelengths are swept, while the generated light is detected and recorded synchronously using a data-acquisition system. 

\paragraph{Optical measurement setup for Supercontinuum measurements.}
The experimental setup is schematically illustrated in Fig.~\ref{fig4}a. 
A mode-locked femtosecond laser (Melon Systems C-Fiber High Power; wavelength 1560 nm, pulse duration 80 fs, repetition rate 100 MHz) is employed as the pump, whose spectrum is shown in Fig. \ref{fig4}b (gray line). 
The pump pulse energy is controlled using neutral-density (ND) filters. 
A half-wave plate (HWP) is used to align the TE polarization. 
Light is coupled in and out of the waveguide using aspheric lenses. 
Optical spectrum analyzer (OSA, Yokogawa AQ6370E, 750 nm - 1600 nm) and spectrometers (SIMTRUM, L200/1000, 300 nm - 850 nm; N/900-2500p, 1500 nm - 2600 nm) are used to measure the output spectra in Fig. \ref{fig1}e and \ref{fig4}b.
The measured spectra are spliced by calibrating the relative spectral overlap between instruments. 

When measuring the RF beatnotes (Fig.\ref{fig1}c, \ref{figs5}), band-pass filters (BPFs, 400 $\pm$ 12.5 nm, 500 $\pm$ 5 nm, 633 $\pm$ 2.5 nm, 900 $\pm$ 12.5 nm, 1310 $\pm$ 6 nm, and 1650 $\pm$ 6 nm) are placed after coupling the light out of the chip to select specific wavelengths. 
The selected signal is detected using a photodetector, and the resulting RF signals are analyzed with an electrical spectrum analyzer (ESA).
The resolution bandwidth and the video bandwidth are both set as 1 kHz.
  
\section*{Acknowledgements}
This research was supported by the Ministry of Education, Singapore, under its Academic Research Fund Tier 1 (FY2025) and Tier 3 (MOET32024-0009), the National Research Foundation, Singapore, under its NRF Fellowship (NRF-NRFF15-2023-0005), and the China Scholarship Council (202406240214).

\section*{Competing interests}
The authors declare no competing interests.

\section*{Contributions}
D.Z., Y.L., and X.S. conceived the idea. Y.L. and X.S. designed the devices. Y.L., X.S., S.S.M., and X.C. fabricated the devices. Y.L., X.S., M.Z., and X.M. performed the measurements of the devices. D.Z., G.D., S.Z., and Q.W. supervised the project. All the authors discussed the results and wrote the manuscript.
\newline

\bibliography{ref_main}

\begin{thebibliography}{10}

\bibitem{voumard2023simulating}
T.~Voumard, M.~Ludwig, T.~Wildi, F.~Ayhan, V.~Brasch, L.~G. Villanueva, and T.~Herr, ``Simulating supercontinua from mixed and cascaded nonlinearities,'' {\em APL photonics}, vol.~8, no.~3, 2023.

\bibitem{wu2022broadband}
X.~Wu, L.~Zhang, Z.~Hao, R.~Zhang, R.~Ma, F.~Bo, G.~Zhang, and J.~Xu, ``Broadband second-harmonic generation in step-chirped periodically poled lithium niobate waveguides,'' {\em Optics Letters}, vol.~47, no.~7, pp.~1574--1577, 2022.

\bibitem{zhang2022second}
H.~Zhang, Q.~Li, H.~Zhu, L.~Cai, and H.~Hu, ``Second harmonic generation by quasi-phase matching in a lithium niobate thin film,'' {\em Optical Materials Express}, vol.~12, no.~6, pp.~2252--2259, 2022.

\bibitem{fang2026broadband}
X.-X. Fang, G.~Shentu, and H.~Lu, ``Broadband quantum photon source in a step-chirped periodically poled lithium niobate waveguide,'' {\em Optics Express}, vol.~34, no.~3, pp.~3759--3767, 2026.

\bibitem{diddams2020optical}
S.~A. Diddams, K.~Vahala, and T.~Udem, ``Optical frequency combs: Coherently uniting the electromagnetic spectrum,'' {\em Science}, vol.~369, no.~6501, p.~eaay3676, 2020.

\bibitem{fan2025spectral1}
W.~Fan, F.~Ayhan, T.~Wildi, M.~Volkov, A.~Seer, M.~Ludwig, T.~Voumard, A.~Brodschelm, V.~Brasch, L.~G. Villanueva, {\em et~al.}, ``Spectral dynamics in broadband frequency combs with overlapping harmonics,'' {\em Physical Review Letters}, vol.~135, no.~21, p.~213801, 2025.

\bibitem{wu2024visible1}
T.-H. Wu, L.~Ledezma, C.~Fredrick, P.~Sekhar, R.~Sekine, Q.~Guo, R.~M. Briggs, A.~Marandi, and S.~A. Diddams, ``Visible-to-ultraviolet frequency comb generation in lithium niobate nanophotonic waveguides,'' {\em Nature Photonics}, vol.~18, no.~3, pp.~218--223, 2024.

\bibitem{leo2014dispersive}
F.~Leo, S.-P. Gorza, J.~Safioui, P.~Kockaert, S.~Coen, U.~Dave, B.~Kuyken, and G.~Roelkens, ``Dispersive wave emission and supercontinuum generation in a silicon wire waveguide pumped around the 1550 nm telecommunication wavelength,'' {\em Optics letters}, vol.~39, no.~12, pp.~3623--3626, 2014.

\bibitem{singh2018octave}
N.~Singh, M.~Xin, D.~Vermeulen, K.~Shtyrkova, N.~Li, P.~T. Callahan, E.~S. Magden, A.~Ruocco, N.~Fahrenkopf, C.~Baiocco, {\em et~al.}, ``Octave-spanning coherent supercontinuum generation in silicon on insulator from 1.06 $\mu$m to beyond 2.4 $\mu$m,'' {\em Light: Science \& Applications}, vol.~7, no.~1, pp.~17131--17131, 2018.

\bibitem{yoon2017coherent}
D.~Yoon~Oh, K.~Y. Yang, C.~Fredrick, G.~Ycas, S.~A. Diddams, and K.~J. Vahala, ``Coherent ultra-violet to near-infrared generation in silica ridge waveguides,'' {\em Nature communications}, vol.~8, no.~1, p.~13922, 2017.

\bibitem{grassani2019mid}
D.~Grassani, E.~Tagkoudi, H.~Guo, C.~Herkommer, F.~Yang, T.~J. Kippenberg, and C.-S. Br{\`e}s, ``Mid infrared gas spectroscopy using efficient fiber laser driven photonic chip-based supercontinuum,'' {\em Nature communications}, vol.~10, no.~1, p.~1553, 2019.

\bibitem{porcel2017two}
M.~A. Porcel, F.~Schepers, J.~P. Epping, T.~Hellwig, M.~Hoekman, R.~G. Heideman, P.~J. van~der Slot, C.~J. Lee, R.~Schmidt, R.~Bratschitsch, {\em et~al.}, ``Two-octave spanning supercontinuum generation in stoichiometric silicon nitride waveguides pumped at telecom wavelengths,'' {\em Optics express}, vol.~25, no.~2, pp.~1542--1554, 2017.

\bibitem{deniel2023visible}
L.~Deniel, M.~A. Guidry, D.~M. Lukin, K.~Y. Yang, J.~Yang, J.~Vu{\v{c}}kovi{\'c}, T.~W. H{\"a}nsch, and N.~Picqu{\'e}, ``Visible to mid-infrared supercontinuum generation in 4h-silicon-carbide nanophotonic waveguides,'' in {\em CLEO: Science and Innovations}, pp.~STh1F--4, Optica Publishing Group, 2023.

\bibitem{yan2026simplified}
H.~Yan, S.~Zhang, A.~Pal, A.~Ghosh, A.~Alabbadi, M.~Kheyri, T.~Bi, Y.~Zhang, I.~Harder, O.~Ohletz, {\em et~al.}, ``Simplified aluminum nitride processing for low-loss integrated photonics and nonlinear optics,'' {\em npj Nanophotonics}, vol.~3, no.~1, p.~13, 2026.

\bibitem{kuyken2020octave}
B.~Kuyken, M.~Billet, F.~Leo, K.~Yvind, and M.~Pu, ``Octave-spanning coherent supercontinuum generation in an algaas-on-insulator waveguide,'' {\em Optics Letters}, vol.~45, no.~3, pp.~603--606, 2020.

\bibitem{fan2024supercontinua}
W.~Fan, M.~Ludwig, I.~Rousseau, I.~Arabadzhiev, B.~Ruhnke, T.~Wildi, and T.~Herr, ``Supercontinua from integrated gallium nitride waveguides,'' {\em Optica}, vol.~11, no.~8, pp.~1175--1181, 2024.

\bibitem{dave2015dispersive}
U.~D. Dave, C.~Ciret, S.-P. Gorza, S.~Combrie, A.~De~Rossi, F.~Raineri, G.~Roelkens, and B.~Kuyken, ``Dispersive-wave-based octave-spanning supercontinuum generation in ingap membrane waveguides on a silicon substrate,'' {\em Optics letters}, vol.~40, no.~15, pp.~3584--3587, 2015.

\bibitem{jung2021tantala}
H.~Jung, S.-P. Yu, D.~R. Carlson, T.~E. Drake, T.~C. Briles, and S.~B. Papp, ``Tantala kerr nonlinear integrated photonics,'' {\em Optica}, vol.~8, no.~6, pp.~811--817, 2021.

\bibitem{singh2020nonlinear}
N.~Singh, H.~M. Mbonde, H.~C. Frankis, E.~Ippen, J.~D. Bradley, and F.~X. K{\"a}rtner, ``Nonlinear silicon photonics on cmos-compatible tellurium oxide,'' {\em Photonics Research}, vol.~8, no.~12, pp.~1904--1909, 2020.

\bibitem{hammani2018octave}
K.~Hammani, L.~Markey, M.~Lamy, B.~Kibler, J.~Arocas, J.~Fatome, A.~Dereux, J.-C. Weeber, and C.~Finot, ``Octave spanning supercontinuum in titanium dioxide waveguides,'' {\em Applied Sciences}, vol.~8, no.~4, p.~543, 2018.

\bibitem{hwang2021supercontinuum}
J.~Hwang, D.-G. Kim, S.~Han, D.~Jeong, Y.-H. Lee, D.-Y. Choi, and H.~Lee, ``Supercontinuum generation in as2s3 waveguides fabricated without direct etching,'' {\em Optics Letters}, vol.~46, no.~10, pp.~2413--2416, 2021.

\bibitem{choi2016nonlinear}
J.~W. Choi, Z.~Han, B.-U. Sohn, G.~F. Chen, C.~Smith, L.~C. Kimerling, K.~A. Richardson, A.~M. Agarwal, and D.~T. Tan, ``Nonlinear characterization of gesbs chalcogenide glass waveguides,'' {\em Scientific Reports}, vol.~6, no.~1, p.~39234, 2016.

\bibitem{wang2025three}
L.~Wang, T.~Tang, H.~Zhu, and J.~Lu, ``Three-octave supercontinuum generation spanning from ultraviolet in lithium tantalate waveguides,'' {\em arXiv preprint arXiv:2512.16350}, 2025.

\bibitem{peng2025three}
L.~Peng, X.~Li, L.~Hong, B.~Chen, Y.~Zhao, X.~Duan, H.~Yu, and Z.~Li, ``Three-octave-spanning supercontinuum generation in z-cut quasi-phase matching thin-film lithium niobate,'' {\em APL Photonics}, vol.~10, no.~7, 2025.

\bibitem{hamrouni2024picojoule}
M.~Hamrouni, M.~Jankowski, A.~Y. Hwang, N.~Flemens, J.~Mishra, C.~Langrock, A.~H. Safavi-Naeini, M.~M. Fejer, and T.~S{\"u}dmeyer, ``Picojoule-level supercontinuum generation in thin-film lithium niobate on sapphire,'' {\em Optics Express}, vol.~32, no.~7, pp.~12004--12011, 2024.

\bibitem{li20262}
M.~Li, Q.~Li, Y.~Chu, Y.~Liang, H.~Guo, X.~Sun, H.~Shi, X.~Zheng, J.~Lin, and Y.~Cheng, ``2.7-octave supercontinuum generation spanning from ultraviolet to near-infrared in thin-film lithium niobate waveguides,'' {\em Advanced Optical Materials}, vol.~14, no.~3, p.~e02245, 2026.

\bibitem{zhou2025quadratic}
S.~Zhou, M.~Shen, R.~Sekine, N.~Englebert, T.~Zacharias, B.~Gutierrez, R.~M. Gray, J.~Widjaja, and A.~Marandi, ``Quadratic supercontinuum generation from uv to mid-ir in lithium niobate nanophotonics,'' {\em arXiv preprint arXiv:2510.18844}, 2025.

\bibitem{jankowski2020ultrabroadband}
M.~Jankowski, C.~Langrock, B.~Desiatov, A.~Marandi, C.~Wang, M.~Zhang, C.~R. Phillips, M.~Lon{\v{c}}ar, and M.~M. Fejer, ``Ultrabroadband nonlinear optics in nanophotonic periodically poled lithium niobate waveguides,'' {\em Optica}, vol.~7, no.~1, pp.~40--46, 2020.

\bibitem{yu2019coherent}
M.~Yu, B.~Desiatov, Y.~Okawachi, A.~L. Gaeta, and M.~Lon{\v{c}}ar, ``Coherent two-octave-spanning supercontinuum generation in lithium-niobate waveguides,'' {\em Optics letters}, vol.~44, no.~5, pp.~1222--1225, 2019.

\bibitem{fan2025spectral}
W.~Fan, F.~Ayhan, T.~Wildi, M.~Volkov, A.~Seer, M.~Ludwig, T.~Voumard, A.~Brodschelm, V.~Brasch, L.~G. Villanueva, {\em et~al.}, ``Spectral dynamics in broadband frequency combs with overlapping harmonics,'' {\em Physical Review Letters}, vol.~135, no.~21, p.~213801, 2025.

\bibitem{lu2019octave}
J.~Lu, J.~B. Surya, X.~Liu, Y.~Xu, and H.~X. Tang, ``Octave-spanning supercontinuum generation in nanoscale lithium niobate waveguides,'' {\em Optics letters}, vol.~44, no.~6, pp.~1492--1495, 2019.

\bibitem{wu2024visible}
T.-H. Wu, L.~Ledezma, C.~Fredrick, P.~Sekhar, R.~Sekine, Q.~Guo, R.~M. Briggs, A.~Marandi, and S.~A. Diddams, ``Visible-to-ultraviolet frequency comb generation in lithium niobate nanophotonic waveguides,'' {\em Nature Photonics}, vol.~18, no.~3, pp.~218--223, 2024.

\bibitem{Tang:25}
Y.~Tang, J.~Qiu, W.~Ding, T.~Ding, X.~Song, T.~Xian, H.~Li, S.~Liu, L.~Yuan, Y.~Zheng, and X.~Chen, ``Lithium niobate micro-waveguides for efficient supercontinuum generation and frequency comb self-referencing,'' {\em Photon. Res.}, vol.~13, pp.~3332--3340, Dec 2025.

\bibitem{10.1063/5.0028776}
M.~Reig~Escalé, F.~Kaufmann, H.~Jiang, D.~Pohl, and R.~Grange, ``Generation of 280 thz-spanning near-ultraviolet light in lithium niobate-on-insulator waveguides with sub-100 pj pulses,'' {\em APL Photonics}, vol.~5, p.~121301, 12 2020.

\bibitem{ayhan2025fabrication}
F.~Ayhan, M.~Ludwig, T.~Herr, V.~Brasch, and L.~G. Villanueva, ``Fabrication of periodically poled lithium niobate waveguides for broadband nonlinear photonics,'' {\em APL Photonics}, vol.~10, no.~1, 2025.

\bibitem{ludwig2025mid}
M.~Ludwig, F.~Ayhan, T.~Voumard, W.~Fan, M.~A. Gaafar, V.~Brasch, L.~G. Villanueva, and T.~Herr, ``Mid-infrared continua via spectral broadening and difference frequency generation in a nanophotonic lithium niobate waveguide,'' {\em arXiv preprint arXiv:2510.23878}, 2025.

\bibitem{tang2025chip}
T.~Tang, S.~Yu, R.~Zhou, J.~Zhang, J.~Lu, G.~Chen, T.~Zhu, and L.~Wang, ``On-chip amplification-free fceo detection and broadband scg in parabolically width-modulated tfln waveguides,'' {\em arXiv preprint arXiv:2509.00394}, 2025.

\bibitem{gao2025tightly}
Y.~Gao, Y.~Sun, I.~Rebolledo-Salgado, R.~Van~Laer, V.~Torres-Company, and J.~Schr{\"o}der, ``Tightly-confined and long z-cut lithium niobate waveguide with ultralow-loss,'' {\em Laser \& Photonics Reviews}, vol.~19, no.~21, p.~e00042, 2025.

\end{thebibliography}


\begin{thebibliography}{60}%
\makeatletter
\providecommand \@ifxundefined [1]{%
 \@ifx{#1\undefined}
}%
\providecommand \@ifnum [1]{%
 \ifnum #1\expandafter \@firstoftwo
 \else \expandafter \@secondoftwo
 \fi
}%
\providecommand \@ifx [1]{%
 \ifx #1\expandafter \@firstoftwo
 \else \expandafter \@secondoftwo
 \fi
}%
\providecommand \natexlab [1]{#1}%
\providecommand \enquote  [1]{``#1''}%
\providecommand \bibnamefont  [1]{#1}%
\providecommand \bibfnamefont [1]{#1}%
\providecommand \citenamefont [1]{#1}%
\providecommand \href@noop [0]{\@secondoftwo}%
\providecommand \href [0]{\begingroup \@sanitize@url \@href}%
\providecommand \@href[1]{\@@startlink{#1}\@@href}%
\providecommand \@@href[1]{\endgroup#1\@@endlink}%
\providecommand \@sanitize@url [0]{\catcode `\\12\catcode `\$12\catcode `\&12\catcode `\#12\catcode `\^12\catcode `\_12\catcode `\%12\relax}%
\providecommand \@@startlink[1]{}%
\providecommand \@@endlink[0]{}%
\providecommand \url  [0]{\begingroup\@sanitize@url \@url }%
\providecommand \@url [1]{\endgroup\@href {#1}{\urlprefix }}%
\providecommand \urlprefix  [0]{URL }%
\providecommand \Eprint [0]{\href }%
\providecommand \doibase [0]{https://doi.org/}%
\providecommand \selectlanguage [0]{\@gobble}%
\providecommand \bibinfo  [0]{\@secondoftwo}%
\providecommand \bibfield  [0]{\@secondoftwo}%
\providecommand \translation [1]{[#1]}%
\providecommand \BibitemOpen [0]{}%
\providecommand \bibitemStop [0]{}%
\providecommand \bibitemNoStop [0]{.\EOS\space}%
\providecommand \EOS [0]{\spacefactor3000\relax}%
\providecommand \BibitemShut  [1]{\csname bibitem#1\endcsname}%
\let\auto@bib@innerbib\@empty
\bibitem [{\citenamefont {Drexler}(2004)}]{drexler2004ultrahigh}%
  \BibitemOpen
  \bibfield  {author} {\bibinfo {author} {\bibfnamefont {W.}~\bibnamefont {Drexler}},\ }\bibfield  {title} {\enquote {\bibinfo {title} {Ultrahigh-resolution optical coherence tomography},}\ }\href@noop {} {\bibfield  {journal} {\bibinfo  {journal} {Journal of biomedical optics}\ }\textbf {\bibinfo {volume} {9}},\ \bibinfo {pages} {47--74} (\bibinfo {year} {2004})}\BibitemShut {NoStop}%
\bibitem [{\citenamefont {Hartl}\ \emph {et~al.}(2001)\citenamefont {Hartl}, \citenamefont {Li}, \citenamefont {Chudoba}, \citenamefont {Ghanta}, \citenamefont {Ko}, \citenamefont {Fujimoto}, \citenamefont {Ranka},\ and\ \citenamefont {Windeler}}]{hartl2001ultrahigh}%
  \BibitemOpen
  \bibfield  {author} {\bibinfo {author} {\bibfnamefont {I.}~\bibnamefont {Hartl}}, \bibinfo {author} {\bibfnamefont {X.}~\bibnamefont {Li}}, \bibinfo {author} {\bibfnamefont {C.}~\bibnamefont {Chudoba}}, \bibinfo {author} {\bibfnamefont {R.}~\bibnamefont {Ghanta}}, \bibinfo {author} {\bibfnamefont {T.}~\bibnamefont {Ko}}, \bibinfo {author} {\bibfnamefont {J.}~\bibnamefont {Fujimoto}}, \bibinfo {author} {\bibfnamefont {J.}~\bibnamefont {Ranka}},\ and\ \bibinfo {author} {\bibfnamefont {R.}~\bibnamefont {Windeler}},\ }\bibfield  {title} {\enquote {\bibinfo {title} {Ultrahigh-resolution optical coherence tomography using continuum generation in an air--silica microstructure optical fiber},}\ }\href@noop {} {\bibfield  {journal} {\bibinfo  {journal} {Optics letters}\ }\textbf {\bibinfo {volume} {26}},\ \bibinfo {pages} {608--610} (\bibinfo {year} {2001})}\BibitemShut {NoStop}%
\bibitem [{\citenamefont {Nishizawa}\ \emph {et~al.}(2004)\citenamefont {Nishizawa}, \citenamefont {Chen}, \citenamefont {Hsiung}, \citenamefont {Ippen},\ and\ \citenamefont {Fujimoto}}]{nishizawa2004real}%
  \BibitemOpen
  \bibfield  {author} {\bibinfo {author} {\bibfnamefont {N.}~\bibnamefont {Nishizawa}}, \bibinfo {author} {\bibfnamefont {Y.}~\bibnamefont {Chen}}, \bibinfo {author} {\bibfnamefont {P.}~\bibnamefont {Hsiung}}, \bibinfo {author} {\bibfnamefont {E.}~\bibnamefont {Ippen}},\ and\ \bibinfo {author} {\bibfnamefont {J.}~\bibnamefont {Fujimoto}},\ }\bibfield  {title} {\enquote {\bibinfo {title} {Real-time, ultrahigh-resolution, optical coherence tomography with an all-fiber, femtosecond fiber laser continuum at 1.5 $\mu$ m},}\ }\href@noop {} {\bibfield  {journal} {\bibinfo  {journal} {Optics letters}\ }\textbf {\bibinfo {volume} {29}},\ \bibinfo {pages} {2846--2848} (\bibinfo {year} {2004})}\BibitemShut {NoStop}%
\bibitem [{\citenamefont {Sotobayashi}, \citenamefont {Chujo},\ and\ \citenamefont {Ozeki}(2001)}]{sotobayashi2001bi}%
  \BibitemOpen
  \bibfield  {author} {\bibinfo {author} {\bibfnamefont {H.}~\bibnamefont {Sotobayashi}}, \bibinfo {author} {\bibfnamefont {W.}~\bibnamefont {Chujo}},\ and\ \bibinfo {author} {\bibfnamefont {T.}~\bibnamefont {Ozeki}},\ }\bibfield  {title} {\enquote {\bibinfo {title} {Bi-directional photonic conversion between 4x10 gbit/s otdm and wdm by optical time-gating wavelength interchange},}\ }in\ \href@noop {} {\emph {\bibinfo {booktitle} {Optical Fiber Communication Conference}}}\ (\bibinfo {organization} {Optica Publishing Group},\ \bibinfo {year} {2001})\ p.\ \bibinfo {pages} {WM5}\BibitemShut {NoStop}%
\bibitem [{\citenamefont {Sotobayashi}\ \emph {et~al.}(2002)\citenamefont {Sotobayashi}, \citenamefont {Chujo}, \citenamefont {Konishi},\ and\ \citenamefont {Ozeki}}]{sotobayashi2002wavelength}%
  \BibitemOpen
  \bibfield  {author} {\bibinfo {author} {\bibfnamefont {H.}~\bibnamefont {Sotobayashi}}, \bibinfo {author} {\bibfnamefont {W.}~\bibnamefont {Chujo}}, \bibinfo {author} {\bibfnamefont {A.}~\bibnamefont {Konishi}},\ and\ \bibinfo {author} {\bibfnamefont {T.}~\bibnamefont {Ozeki}},\ }\bibfield  {title} {\enquote {\bibinfo {title} {Wavelength-band generation and transmission of 3.24-tbit/s (81-channel wdm$\times$ 40-gbit/s) carrier-suppressed return-to-zero format by use of a single supercontinuum source for frequency standardization},}\ }\href@noop {} {\bibfield  {journal} {\bibinfo  {journal} {Journal of the Optical Society of America B}\ }\textbf {\bibinfo {volume} {19}},\ \bibinfo {pages} {2803--2809} (\bibinfo {year} {2002})}\BibitemShut {NoStop}%
\bibitem [{\citenamefont {Hu}\ \emph {et~al.}(2018)\citenamefont {Hu}, \citenamefont {Da~Ros}, \citenamefont {Pu}, \citenamefont {Ye}, \citenamefont {Ingerslev}, \citenamefont {Porto~da Silva}, \citenamefont {Nooruzzaman}, \citenamefont {Amma}, \citenamefont {Sasaki}, \citenamefont {Mizuno} \emph {et~al.}}]{hu2018single}%
  \BibitemOpen
  \bibfield  {author} {\bibinfo {author} {\bibfnamefont {H.}~\bibnamefont {Hu}}, \bibinfo {author} {\bibfnamefont {F.}~\bibnamefont {Da~Ros}}, \bibinfo {author} {\bibfnamefont {M.}~\bibnamefont {Pu}}, \bibinfo {author} {\bibfnamefont {F.}~\bibnamefont {Ye}}, \bibinfo {author} {\bibfnamefont {K.}~\bibnamefont {Ingerslev}}, \bibinfo {author} {\bibfnamefont {E.}~\bibnamefont {Porto~da Silva}}, \bibinfo {author} {\bibfnamefont {M.}~\bibnamefont {Nooruzzaman}}, \bibinfo {author} {\bibfnamefont {Y.}~\bibnamefont {Amma}}, \bibinfo {author} {\bibfnamefont {Y.}~\bibnamefont {Sasaki}}, \bibinfo {author} {\bibfnamefont {T.}~\bibnamefont {Mizuno}}, \emph {et~al.},\ }\bibfield  {title} {\enquote {\bibinfo {title} {Single-source chip-based frequency comb enabling extreme parallel data transmission},}\ }\href@noop {} {\bibfield  {journal} {\bibinfo  {journal} {Nature Photonics}\ }\textbf {\bibinfo {volume} {12}},\ \bibinfo {pages} {469--473} (\bibinfo {year} {2018})}\BibitemShut {NoStop}%
\bibitem [{\citenamefont {Diddams}, \citenamefont {Hollberg},\ and\ \citenamefont {Mbele}(2007)}]{diddams2007molecular}%
  \BibitemOpen
  \bibfield  {author} {\bibinfo {author} {\bibfnamefont {S.~A.}\ \bibnamefont {Diddams}}, \bibinfo {author} {\bibfnamefont {L.}~\bibnamefont {Hollberg}},\ and\ \bibinfo {author} {\bibfnamefont {V.}~\bibnamefont {Mbele}},\ }\bibfield  {title} {\enquote {\bibinfo {title} {Molecular fingerprinting with the resolved modes of a femtosecond laser frequency comb},}\ }\href@noop {} {\bibfield  {journal} {\bibinfo  {journal} {Nature}\ }\textbf {\bibinfo {volume} {445}},\ \bibinfo {pages} {627--630} (\bibinfo {year} {2007})}\BibitemShut {NoStop}%
\bibitem [{\citenamefont {Timmers}\ \emph {et~al.}(2018)\citenamefont {Timmers}, \citenamefont {Kowligy}, \citenamefont {Lind}, \citenamefont {Cruz}, \citenamefont {Nader}, \citenamefont {Silfies}, \citenamefont {Ycas}, \citenamefont {Allison}, \citenamefont {Schunemann}, \citenamefont {Papp} \emph {et~al.}}]{timmers2018molecular}%
  \BibitemOpen
  \bibfield  {author} {\bibinfo {author} {\bibfnamefont {H.}~\bibnamefont {Timmers}}, \bibinfo {author} {\bibfnamefont {A.}~\bibnamefont {Kowligy}}, \bibinfo {author} {\bibfnamefont {A.}~\bibnamefont {Lind}}, \bibinfo {author} {\bibfnamefont {F.~C.}\ \bibnamefont {Cruz}}, \bibinfo {author} {\bibfnamefont {N.}~\bibnamefont {Nader}}, \bibinfo {author} {\bibfnamefont {M.}~\bibnamefont {Silfies}}, \bibinfo {author} {\bibfnamefont {G.}~\bibnamefont {Ycas}}, \bibinfo {author} {\bibfnamefont {T.~K.}\ \bibnamefont {Allison}}, \bibinfo {author} {\bibfnamefont {P.~G.}\ \bibnamefont {Schunemann}}, \bibinfo {author} {\bibfnamefont {S.~B.}\ \bibnamefont {Papp}}, \emph {et~al.},\ }\bibfield  {title} {\enquote {\bibinfo {title} {Molecular fingerprinting with bright, broadband infrared frequency combs},}\ }\href@noop {} {\bibfield  {journal} {\bibinfo  {journal} {Optica}\ }\textbf {\bibinfo {volume} {5}},\ \bibinfo {pages} {727--732} (\bibinfo {year} {2018})}\BibitemShut {NoStop}%
\bibitem [{\citenamefont {Min}\ \emph {et~al.}(2011)\citenamefont {Min}, \citenamefont {Freudiger}, \citenamefont {Lu},\ and\ \citenamefont {Xie}}]{min2011coherent}%
  \BibitemOpen
  \bibfield  {author} {\bibinfo {author} {\bibfnamefont {W.}~\bibnamefont {Min}}, \bibinfo {author} {\bibfnamefont {C.~W.}\ \bibnamefont {Freudiger}}, \bibinfo {author} {\bibfnamefont {S.}~\bibnamefont {Lu}},\ and\ \bibinfo {author} {\bibfnamefont {X.~S.}\ \bibnamefont {Xie}},\ }\bibfield  {title} {\enquote {\bibinfo {title} {Coherent nonlinear optical imaging: beyond fluorescence microscopy},}\ }\href@noop {} {\bibfield  {journal} {\bibinfo  {journal} {Annual review of physical chemistry}\ }\textbf {\bibinfo {volume} {62}},\ \bibinfo {pages} {507--530} (\bibinfo {year} {2011})}\BibitemShut {NoStop}%
\bibitem [{\citenamefont {Du}\ \emph {et~al.}(2019)\citenamefont {Du}, \citenamefont {Zhang}, \citenamefont {Li}, \citenamefont {Gao},\ and\ \citenamefont {Tong}}]{du2019mid}%
  \BibitemOpen
  \bibfield  {author} {\bibinfo {author} {\bibfnamefont {Z.}~\bibnamefont {Du}}, \bibinfo {author} {\bibfnamefont {S.}~\bibnamefont {Zhang}}, \bibinfo {author} {\bibfnamefont {J.}~\bibnamefont {Li}}, \bibinfo {author} {\bibfnamefont {N.}~\bibnamefont {Gao}},\ and\ \bibinfo {author} {\bibfnamefont {K.}~\bibnamefont {Tong}},\ }\bibfield  {title} {\enquote {\bibinfo {title} {Mid-infrared tunable laser-based broadband fingerprint absorption spectroscopy for trace gas sensing: A review},}\ }\href@noop {} {\bibfield  {journal} {\bibinfo  {journal} {Applied sciences}\ }\textbf {\bibinfo {volume} {9}},\ \bibinfo {pages} {338} (\bibinfo {year} {2019})}\BibitemShut {NoStop}%
\bibitem [{\citenamefont {Gaeta}, \citenamefont {Lipson},\ and\ \citenamefont {Kippenberg}(2019)}]{gaeta2019photonic}%
  \BibitemOpen
  \bibfield  {author} {\bibinfo {author} {\bibfnamefont {A.~L.}\ \bibnamefont {Gaeta}}, \bibinfo {author} {\bibfnamefont {M.}~\bibnamefont {Lipson}},\ and\ \bibinfo {author} {\bibfnamefont {T.~J.}\ \bibnamefont {Kippenberg}},\ }\bibfield  {title} {\enquote {\bibinfo {title} {Photonic-chip-based frequency combs},}\ }\href@noop {} {\bibfield  {journal} {\bibinfo  {journal} {nature photonics}\ }\textbf {\bibinfo {volume} {13}},\ \bibinfo {pages} {158--169} (\bibinfo {year} {2019})}\BibitemShut {NoStop}%
\bibitem [{\citenamefont {Dudley}, \citenamefont {Genty},\ and\ \citenamefont {Coen}(2006)}]{dudley2006supercontinuum}%
  \BibitemOpen
  \bibfield  {author} {\bibinfo {author} {\bibfnamefont {J.~M.}\ \bibnamefont {Dudley}}, \bibinfo {author} {\bibfnamefont {G.}~\bibnamefont {Genty}},\ and\ \bibinfo {author} {\bibfnamefont {S.}~\bibnamefont {Coen}},\ }\bibfield  {title} {\enquote {\bibinfo {title} {Supercontinuum generation in photonic crystal fiber},}\ }\href@noop {} {\bibfield  {journal} {\bibinfo  {journal} {Reviews of modern physics}\ }\textbf {\bibinfo {volume} {78}},\ \bibinfo {pages} {1135--1184} (\bibinfo {year} {2006})}\BibitemShut {NoStop}%
\bibitem [{\citenamefont {Birks}, \citenamefont {Wadsworth},\ and\ \citenamefont {Russell}(2000)}]{birks2000supercontinuum}%
  \BibitemOpen
  \bibfield  {author} {\bibinfo {author} {\bibfnamefont {T.}~\bibnamefont {Birks}}, \bibinfo {author} {\bibfnamefont {W.}~\bibnamefont {Wadsworth}},\ and\ \bibinfo {author} {\bibfnamefont {P.~S.~J.}\ \bibnamefont {Russell}},\ }\bibfield  {title} {\enquote {\bibinfo {title} {Supercontinuum generation in tapered fibers},}\ }\href@noop {} {\bibfield  {journal} {\bibinfo  {journal} {Optics letters}\ }\textbf {\bibinfo {volume} {25}},\ \bibinfo {pages} {1415--1417} (\bibinfo {year} {2000})}\BibitemShut {NoStop}%
\bibitem [{\citenamefont {Hu}\ \emph {et~al.}(2010)\citenamefont {Hu}, \citenamefont {Menyuk}, \citenamefont {Shaw}, \citenamefont {Sanghera},\ and\ \citenamefont {Aggarwal}}]{hu2010maximizing}%
  \BibitemOpen
  \bibfield  {author} {\bibinfo {author} {\bibfnamefont {J.}~\bibnamefont {Hu}}, \bibinfo {author} {\bibfnamefont {C.~R.}\ \bibnamefont {Menyuk}}, \bibinfo {author} {\bibfnamefont {L.~B.}\ \bibnamefont {Shaw}}, \bibinfo {author} {\bibfnamefont {J.~S.}\ \bibnamefont {Sanghera}},\ and\ \bibinfo {author} {\bibfnamefont {I.~D.}\ \bibnamefont {Aggarwal}},\ }\bibfield  {title} {\enquote {\bibinfo {title} {Maximizing the bandwidth of supercontinuum generation in as2se3 chalcogenide fibers},}\ }\href@noop {} {\bibfield  {journal} {\bibinfo  {journal} {Optics express}\ }\textbf {\bibinfo {volume} {18}},\ \bibinfo {pages} {6722--6739} (\bibinfo {year} {2010})}\BibitemShut {NoStop}%
\bibitem [{\citenamefont {Johnson}\ \emph {et~al.}(2015)\citenamefont {Johnson}, \citenamefont {Mayer}, \citenamefont {Klenner}, \citenamefont {Luke}, \citenamefont {Lamb}, \citenamefont {Lamont}, \citenamefont {Joshi}, \citenamefont {Okawachi}, \citenamefont {Wise}, \citenamefont {Lipson} \emph {et~al.}}]{johnson2015octave}%
  \BibitemOpen
  \bibfield  {author} {\bibinfo {author} {\bibfnamefont {A.~R.}\ \bibnamefont {Johnson}}, \bibinfo {author} {\bibfnamefont {A.~S.}\ \bibnamefont {Mayer}}, \bibinfo {author} {\bibfnamefont {A.}~\bibnamefont {Klenner}}, \bibinfo {author} {\bibfnamefont {K.}~\bibnamefont {Luke}}, \bibinfo {author} {\bibfnamefont {E.~S.}\ \bibnamefont {Lamb}}, \bibinfo {author} {\bibfnamefont {M.~R.}\ \bibnamefont {Lamont}}, \bibinfo {author} {\bibfnamefont {C.}~\bibnamefont {Joshi}}, \bibinfo {author} {\bibfnamefont {Y.}~\bibnamefont {Okawachi}}, \bibinfo {author} {\bibfnamefont {F.~W.}\ \bibnamefont {Wise}}, \bibinfo {author} {\bibfnamefont {M.}~\bibnamefont {Lipson}}, \emph {et~al.},\ }\bibfield  {title} {\enquote {\bibinfo {title} {Octave-spanning coherent supercontinuum generation in a silicon nitride waveguide},}\ }\href@noop {} {\bibfield  {journal} {\bibinfo  {journal} {Optics letters}\ }\textbf {\bibinfo {volume} {40}},\ \bibinfo {pages} {5117--5120} (\bibinfo {year} {2015})}\BibitemShut {NoStop}%
\bibitem [{\citenamefont {Yan}\ \emph {et~al.}(2026)\citenamefont {Yan}, \citenamefont {Zhang}, \citenamefont {Pal}, \citenamefont {Ghosh}, \citenamefont {Alabbadi}, \citenamefont {Kheyri}, \citenamefont {Bi}, \citenamefont {Zhang}, \citenamefont {Harder}, \citenamefont {Ohletz} \emph {et~al.}}]{yan2026simplified}%
  \BibitemOpen
  \bibfield  {author} {\bibinfo {author} {\bibfnamefont {H.}~\bibnamefont {Yan}}, \bibinfo {author} {\bibfnamefont {S.}~\bibnamefont {Zhang}}, \bibinfo {author} {\bibfnamefont {A.}~\bibnamefont {Pal}}, \bibinfo {author} {\bibfnamefont {A.}~\bibnamefont {Ghosh}}, \bibinfo {author} {\bibfnamefont {A.}~\bibnamefont {Alabbadi}}, \bibinfo {author} {\bibfnamefont {M.}~\bibnamefont {Kheyri}}, \bibinfo {author} {\bibfnamefont {T.}~\bibnamefont {Bi}}, \bibinfo {author} {\bibfnamefont {Y.}~\bibnamefont {Zhang}}, \bibinfo {author} {\bibfnamefont {I.}~\bibnamefont {Harder}}, \bibinfo {author} {\bibfnamefont {O.}~\bibnamefont {Ohletz}}, \emph {et~al.},\ }\bibfield  {title} {\enquote {\bibinfo {title} {Simplified aluminum nitride processing for low-loss integrated photonics and nonlinear optics},}\ }\href@noop {} {\bibfield  {journal} {\bibinfo  {journal} {npj Nanophotonics}\ }\textbf {\bibinfo {volume} {3}},\ \bibinfo {pages} {13} (\bibinfo {year} {2026})}\BibitemShut {NoStop}%
\bibitem [{\citenamefont {Psaila}\ \emph {et~al.}(2007)\citenamefont {Psaila}, \citenamefont {Thomson}, \citenamefont {Bookey}, \citenamefont {Shen}, \citenamefont {Chiodo}, \citenamefont {Osellame}, \citenamefont {Cerullo}, \citenamefont {Jha},\ and\ \citenamefont {Kar}}]{psaila2007supercontinuum}%
  \BibitemOpen
  \bibfield  {author} {\bibinfo {author} {\bibfnamefont {N.~D.}\ \bibnamefont {Psaila}}, \bibinfo {author} {\bibfnamefont {R.~R.}\ \bibnamefont {Thomson}}, \bibinfo {author} {\bibfnamefont {H.~T.}\ \bibnamefont {Bookey}}, \bibinfo {author} {\bibfnamefont {S.}~\bibnamefont {Shen}}, \bibinfo {author} {\bibfnamefont {N.}~\bibnamefont {Chiodo}}, \bibinfo {author} {\bibfnamefont {R.}~\bibnamefont {Osellame}}, \bibinfo {author} {\bibfnamefont {G.}~\bibnamefont {Cerullo}}, \bibinfo {author} {\bibfnamefont {A.}~\bibnamefont {Jha}},\ and\ \bibinfo {author} {\bibfnamefont {A.~K.}\ \bibnamefont {Kar}},\ }\bibfield  {title} {\enquote {\bibinfo {title} {Supercontinuum generation in an ultrafast laser inscribed chalcogenide glass waveguide},}\ }\href@noop {} {\bibfield  {journal} {\bibinfo  {journal} {Optics express}\ }\textbf {\bibinfo {volume} {15}},\ \bibinfo {pages} {15776--15781} (\bibinfo {year} {2007})}\BibitemShut {NoStop}%
\bibitem [{\citenamefont {Li}\ \emph {et~al.}(2024)\citenamefont {Li}, \citenamefont {Wang}, \citenamefont {Afridi}, \citenamefont {Lu}, \citenamefont {Shi}, \citenamefont {Sun}, \citenamefont {Ou},\ and\ \citenamefont {Li}}]{li2024efficient}%
  \BibitemOpen
  \bibfield  {author} {\bibinfo {author} {\bibfnamefont {J.}~\bibnamefont {Li}}, \bibinfo {author} {\bibfnamefont {R.}~\bibnamefont {Wang}}, \bibinfo {author} {\bibfnamefont {A.~A.}\ \bibnamefont {Afridi}}, \bibinfo {author} {\bibfnamefont {Y.}~\bibnamefont {Lu}}, \bibinfo {author} {\bibfnamefont {X.}~\bibnamefont {Shi}}, \bibinfo {author} {\bibfnamefont {W.}~\bibnamefont {Sun}}, \bibinfo {author} {\bibfnamefont {H.}~\bibnamefont {Ou}},\ and\ \bibinfo {author} {\bibfnamefont {Q.}~\bibnamefont {Li}},\ }\bibfield  {title} {\enquote {\bibinfo {title} {Efficient raman lasing and raman--kerr interaction in an integrated silicon carbide platform},}\ }\href@noop {} {\bibfield  {journal} {\bibinfo  {journal} {ACS photonics}\ }\textbf {\bibinfo {volume} {11}},\ \bibinfo {pages} {795--800} (\bibinfo {year} {2024})}\BibitemShut {NoStop}%
\bibitem [{\citenamefont {Afridi}\ \emph {et~al.}(2024)\citenamefont {Afridi}, \citenamefont {Lu}, \citenamefont {Shi}, \citenamefont {Wang}, \citenamefont {Li}, \citenamefont {Li},\ and\ \citenamefont {Ou}}]{afridi20244h}%
  \BibitemOpen
  \bibfield  {author} {\bibinfo {author} {\bibfnamefont {A.~A.}\ \bibnamefont {Afridi}}, \bibinfo {author} {\bibfnamefont {Y.}~\bibnamefont {Lu}}, \bibinfo {author} {\bibfnamefont {X.}~\bibnamefont {Shi}}, \bibinfo {author} {\bibfnamefont {R.}~\bibnamefont {Wang}}, \bibinfo {author} {\bibfnamefont {J.}~\bibnamefont {Li}}, \bibinfo {author} {\bibfnamefont {Q.}~\bibnamefont {Li}},\ and\ \bibinfo {author} {\bibfnamefont {H.}~\bibnamefont {Ou}},\ }\bibfield  {title} {\enquote {\bibinfo {title} {4h--sic microring resonators—opportunities for nonlinear integrated optics},}\ }\href@noop {} {\bibfield  {journal} {\bibinfo  {journal} {Applied Physics Letters}\ }\textbf {\bibinfo {volume} {124}} (\bibinfo {year} {2024})}\BibitemShut {NoStop}%
\bibitem [{\citenamefont {Zhu}\ \emph {et~al.}(2021)\citenamefont {Zhu}, \citenamefont {Shao}, \citenamefont {Yu}, \citenamefont {Cheng}, \citenamefont {Desiatov}, \citenamefont {Xin}, \citenamefont {Hu}, \citenamefont {Holzgrafe}, \citenamefont {Ghosh}, \citenamefont {Shams-Ansari} \emph {et~al.}}]{zhu2021integrated}%
  \BibitemOpen
  \bibfield  {author} {\bibinfo {author} {\bibfnamefont {D.}~\bibnamefont {Zhu}}, \bibinfo {author} {\bibfnamefont {L.}~\bibnamefont {Shao}}, \bibinfo {author} {\bibfnamefont {M.}~\bibnamefont {Yu}}, \bibinfo {author} {\bibfnamefont {R.}~\bibnamefont {Cheng}}, \bibinfo {author} {\bibfnamefont {B.}~\bibnamefont {Desiatov}}, \bibinfo {author} {\bibfnamefont {C.}~\bibnamefont {Xin}}, \bibinfo {author} {\bibfnamefont {Y.}~\bibnamefont {Hu}}, \bibinfo {author} {\bibfnamefont {J.}~\bibnamefont {Holzgrafe}}, \bibinfo {author} {\bibfnamefont {S.}~\bibnamefont {Ghosh}}, \bibinfo {author} {\bibfnamefont {A.}~\bibnamefont {Shams-Ansari}}, \emph {et~al.},\ }\bibfield  {title} {\enquote {\bibinfo {title} {Integrated photonics on thin-film lithium niobate},}\ }\href@noop {} {\bibfield  {journal} {\bibinfo  {journal} {Advances in Optics and Photonics}\ }\textbf {\bibinfo {volume} {13}},\ \bibinfo {pages} {242--352} (\bibinfo {year} {2021})}\BibitemShut {NoStop}%
\bibitem [{\citenamefont {Peng}\ \emph {et~al.}(2025)\citenamefont {Peng}, \citenamefont {Li}, \citenamefont {Hong}, \citenamefont {Chen}, \citenamefont {Zhao}, \citenamefont {Duan}, \citenamefont {Yu},\ and\ \citenamefont {Li}}]{peng2025three}%
  \BibitemOpen
  \bibfield  {author} {\bibinfo {author} {\bibfnamefont {L.}~\bibnamefont {Peng}}, \bibinfo {author} {\bibfnamefont {X.}~\bibnamefont {Li}}, \bibinfo {author} {\bibfnamefont {L.}~\bibnamefont {Hong}}, \bibinfo {author} {\bibfnamefont {B.}~\bibnamefont {Chen}}, \bibinfo {author} {\bibfnamefont {Y.}~\bibnamefont {Zhao}}, \bibinfo {author} {\bibfnamefont {X.}~\bibnamefont {Duan}}, \bibinfo {author} {\bibfnamefont {H.}~\bibnamefont {Yu}},\ and\ \bibinfo {author} {\bibfnamefont {Z.}~\bibnamefont {Li}},\ }\bibfield  {title} {\enquote {\bibinfo {title} {Three-octave-spanning supercontinuum generation in z-cut quasi-phase matching thin-film lithium niobate},}\ }\href@noop {} {\bibfield  {journal} {\bibinfo  {journal} {APL Photonics}\ }\textbf {\bibinfo {volume} {10}} (\bibinfo {year} {2025})}\BibitemShut {NoStop}%
\bibitem [{\citenamefont {Hamrouni}\ \emph {et~al.}(2024)\citenamefont {Hamrouni}, \citenamefont {Jankowski}, \citenamefont {Hwang}, \citenamefont {Flemens}, \citenamefont {Mishra}, \citenamefont {Langrock}, \citenamefont {Safavi-Naeini}, \citenamefont {Fejer},\ and\ \citenamefont {S{\"u}dmeyer}}]{hamrouni2024picojoule}%
  \BibitemOpen
  \bibfield  {author} {\bibinfo {author} {\bibfnamefont {M.}~\bibnamefont {Hamrouni}}, \bibinfo {author} {\bibfnamefont {M.}~\bibnamefont {Jankowski}}, \bibinfo {author} {\bibfnamefont {A.~Y.}\ \bibnamefont {Hwang}}, \bibinfo {author} {\bibfnamefont {N.}~\bibnamefont {Flemens}}, \bibinfo {author} {\bibfnamefont {J.}~\bibnamefont {Mishra}}, \bibinfo {author} {\bibfnamefont {C.}~\bibnamefont {Langrock}}, \bibinfo {author} {\bibfnamefont {A.~H.}\ \bibnamefont {Safavi-Naeini}}, \bibinfo {author} {\bibfnamefont {M.~M.}\ \bibnamefont {Fejer}},\ and\ \bibinfo {author} {\bibfnamefont {T.}~\bibnamefont {S{\"u}dmeyer}},\ }\bibfield  {title} {\enquote {\bibinfo {title} {Picojoule-level supercontinuum generation in thin-film lithium niobate on sapphire},}\ }\href@noop {} {\bibfield  {journal} {\bibinfo  {journal} {Optics Express}\ }\textbf {\bibinfo {volume} {32}},\ \bibinfo {pages} {12004--12011} (\bibinfo {year} {2024})}\BibitemShut {NoStop}%
\bibitem [{\citenamefont {Li}\ \emph {et~al.}(2026)\citenamefont {Li}, \citenamefont {Li}, \citenamefont {Chu}, \citenamefont {Liang}, \citenamefont {Guo}, \citenamefont {Sun}, \citenamefont {Shi}, \citenamefont {Zheng}, \citenamefont {Lin},\ and\ \citenamefont {Cheng}}]{li20262}%
  \BibitemOpen
  \bibfield  {author} {\bibinfo {author} {\bibfnamefont {M.}~\bibnamefont {Li}}, \bibinfo {author} {\bibfnamefont {Q.}~\bibnamefont {Li}}, \bibinfo {author} {\bibfnamefont {Y.}~\bibnamefont {Chu}}, \bibinfo {author} {\bibfnamefont {Y.}~\bibnamefont {Liang}}, \bibinfo {author} {\bibfnamefont {H.}~\bibnamefont {Guo}}, \bibinfo {author} {\bibfnamefont {X.}~\bibnamefont {Sun}}, \bibinfo {author} {\bibfnamefont {H.}~\bibnamefont {Shi}}, \bibinfo {author} {\bibfnamefont {X.}~\bibnamefont {Zheng}}, \bibinfo {author} {\bibfnamefont {J.}~\bibnamefont {Lin}},\ and\ \bibinfo {author} {\bibfnamefont {Y.}~\bibnamefont {Cheng}},\ }\bibfield  {title} {\enquote {\bibinfo {title} {2.7-octave supercontinuum generation spanning from ultraviolet to near-infrared in thin-film lithium niobate waveguides},}\ }\href@noop {} {\bibfield  {journal} {\bibinfo  {journal} {Advanced Optical Materials}\ }\textbf {\bibinfo {volume} {14}},\ \bibinfo {pages} {e02245} (\bibinfo {year} {2026})}\BibitemShut {NoStop}%
\bibitem [{\citenamefont {Zhou}\ \emph {et~al.}(2025)\citenamefont {Zhou}, \citenamefont {Shen}, \citenamefont {Sekine}, \citenamefont {Englebert}, \citenamefont {Zacharias}, \citenamefont {Gutierrez}, \citenamefont {Gray}, \citenamefont {Widjaja},\ and\ \citenamefont {Marandi}}]{zhou2025quadratic}%
  \BibitemOpen
  \bibfield  {author} {\bibinfo {author} {\bibfnamefont {S.}~\bibnamefont {Zhou}}, \bibinfo {author} {\bibfnamefont {M.}~\bibnamefont {Shen}}, \bibinfo {author} {\bibfnamefont {R.}~\bibnamefont {Sekine}}, \bibinfo {author} {\bibfnamefont {N.}~\bibnamefont {Englebert}}, \bibinfo {author} {\bibfnamefont {T.}~\bibnamefont {Zacharias}}, \bibinfo {author} {\bibfnamefont {B.}~\bibnamefont {Gutierrez}}, \bibinfo {author} {\bibfnamefont {R.~M.}\ \bibnamefont {Gray}}, \bibinfo {author} {\bibfnamefont {J.}~\bibnamefont {Widjaja}},\ and\ \bibinfo {author} {\bibfnamefont {A.}~\bibnamefont {Marandi}},\ }\bibfield  {title} {\enquote {\bibinfo {title} {Quadratic supercontinuum generation from uv to mid-ir in lithium niobate nanophotonics},}\ }\href@noop {} {\bibfield  {journal} {\bibinfo  {journal} {arXiv preprint arXiv:2510.18844}\ } (\bibinfo {year} {2025})}\BibitemShut {NoStop}%
\bibitem [{\citenamefont {Jankowski}\ \emph {et~al.}(2020)\citenamefont {Jankowski}, \citenamefont {Langrock}, \citenamefont {Desiatov}, \citenamefont {Marandi}, \citenamefont {Wang}, \citenamefont {Zhang}, \citenamefont {Phillips}, \citenamefont {Lon{\v{c}}ar},\ and\ \citenamefont {Fejer}}]{jankowski2020ultrabroadband}%
  \BibitemOpen
  \bibfield  {author} {\bibinfo {author} {\bibfnamefont {M.}~\bibnamefont {Jankowski}}, \bibinfo {author} {\bibfnamefont {C.}~\bibnamefont {Langrock}}, \bibinfo {author} {\bibfnamefont {B.}~\bibnamefont {Desiatov}}, \bibinfo {author} {\bibfnamefont {A.}~\bibnamefont {Marandi}}, \bibinfo {author} {\bibfnamefont {C.}~\bibnamefont {Wang}}, \bibinfo {author} {\bibfnamefont {M.}~\bibnamefont {Zhang}}, \bibinfo {author} {\bibfnamefont {C.~R.}\ \bibnamefont {Phillips}}, \bibinfo {author} {\bibfnamefont {M.}~\bibnamefont {Lon{\v{c}}ar}},\ and\ \bibinfo {author} {\bibfnamefont {M.~M.}\ \bibnamefont {Fejer}},\ }\bibfield  {title} {\enquote {\bibinfo {title} {Ultrabroadband nonlinear optics in nanophotonic periodically poled lithium niobate waveguides},}\ }\href@noop {} {\bibfield  {journal} {\bibinfo  {journal} {Optica}\ }\textbf {\bibinfo {volume} {7}},\ \bibinfo {pages} {40--46} (\bibinfo {year} {2020})}\BibitemShut {NoStop}%
\bibitem [{\citenamefont {Yu}\ \emph {et~al.}(2019)\citenamefont {Yu}, \citenamefont {Desiatov}, \citenamefont {Okawachi}, \citenamefont {Gaeta},\ and\ \citenamefont {Lon{\v{c}}ar}}]{yu2019coherent}%
  \BibitemOpen
  \bibfield  {author} {\bibinfo {author} {\bibfnamefont {M.}~\bibnamefont {Yu}}, \bibinfo {author} {\bibfnamefont {B.}~\bibnamefont {Desiatov}}, \bibinfo {author} {\bibfnamefont {Y.}~\bibnamefont {Okawachi}}, \bibinfo {author} {\bibfnamefont {A.~L.}\ \bibnamefont {Gaeta}},\ and\ \bibinfo {author} {\bibfnamefont {M.}~\bibnamefont {Lon{\v{c}}ar}},\ }\bibfield  {title} {\enquote {\bibinfo {title} {Coherent two-octave-spanning supercontinuum generation in lithium-niobate waveguides},}\ }\href@noop {} {\bibfield  {journal} {\bibinfo  {journal} {Optics letters}\ }\textbf {\bibinfo {volume} {44}},\ \bibinfo {pages} {1222--1225} (\bibinfo {year} {2019})}\BibitemShut {NoStop}%
\bibitem [{\citenamefont {Fan}\ \emph {et~al.}(2025)\citenamefont {Fan}, \citenamefont {Ayhan}, \citenamefont {Wildi}, \citenamefont {Volkov}, \citenamefont {Seer}, \citenamefont {Ludwig}, \citenamefont {Voumard}, \citenamefont {Brodschelm}, \citenamefont {Brasch}, \citenamefont {Villanueva} \emph {et~al.}}]{fan2025spectral}%
  \BibitemOpen
  \bibfield  {author} {\bibinfo {author} {\bibfnamefont {W.}~\bibnamefont {Fan}}, \bibinfo {author} {\bibfnamefont {F.}~\bibnamefont {Ayhan}}, \bibinfo {author} {\bibfnamefont {T.}~\bibnamefont {Wildi}}, \bibinfo {author} {\bibfnamefont {M.}~\bibnamefont {Volkov}}, \bibinfo {author} {\bibfnamefont {A.}~\bibnamefont {Seer}}, \bibinfo {author} {\bibfnamefont {M.}~\bibnamefont {Ludwig}}, \bibinfo {author} {\bibfnamefont {T.}~\bibnamefont {Voumard}}, \bibinfo {author} {\bibfnamefont {A.}~\bibnamefont {Brodschelm}}, \bibinfo {author} {\bibfnamefont {V.}~\bibnamefont {Brasch}}, \bibinfo {author} {\bibfnamefont {L.~G.}\ \bibnamefont {Villanueva}}, \emph {et~al.},\ }\bibfield  {title} {\enquote {\bibinfo {title} {Spectral dynamics in broadband frequency combs with overlapping harmonics},}\ }\href@noop {} {\bibfield  {journal} {\bibinfo  {journal} {Physical Review Letters}\ }\textbf {\bibinfo {volume} {135}},\ \bibinfo {pages} {213801} (\bibinfo {year} {2025})}\BibitemShut {NoStop}%
\bibitem [{\citenamefont {Lu}\ \emph {et~al.}(2019)\citenamefont {Lu}, \citenamefont {Surya}, \citenamefont {Liu}, \citenamefont {Xu},\ and\ \citenamefont {Tang}}]{lu2019octave}%
  \BibitemOpen
  \bibfield  {author} {\bibinfo {author} {\bibfnamefont {J.}~\bibnamefont {Lu}}, \bibinfo {author} {\bibfnamefont {J.~B.}\ \bibnamefont {Surya}}, \bibinfo {author} {\bibfnamefont {X.}~\bibnamefont {Liu}}, \bibinfo {author} {\bibfnamefont {Y.}~\bibnamefont {Xu}},\ and\ \bibinfo {author} {\bibfnamefont {H.~X.}\ \bibnamefont {Tang}},\ }\bibfield  {title} {\enquote {\bibinfo {title} {Octave-spanning supercontinuum generation in nanoscale lithium niobate waveguides},}\ }\href@noop {} {\bibfield  {journal} {\bibinfo  {journal} {Optics letters}\ }\textbf {\bibinfo {volume} {44}},\ \bibinfo {pages} {1492--1495} (\bibinfo {year} {2019})}\BibitemShut {NoStop}%
\bibitem [{\citenamefont {Wu}\ \emph {et~al.}(2024)\citenamefont {Wu}, \citenamefont {Ledezma}, \citenamefont {Fredrick}, \citenamefont {Sekhar}, \citenamefont {Sekine}, \citenamefont {Guo}, \citenamefont {Briggs}, \citenamefont {Marandi},\ and\ \citenamefont {Diddams}}]{wu2024visible}%
  \BibitemOpen
  \bibfield  {author} {\bibinfo {author} {\bibfnamefont {T.-H.}\ \bibnamefont {Wu}}, \bibinfo {author} {\bibfnamefont {L.}~\bibnamefont {Ledezma}}, \bibinfo {author} {\bibfnamefont {C.}~\bibnamefont {Fredrick}}, \bibinfo {author} {\bibfnamefont {P.}~\bibnamefont {Sekhar}}, \bibinfo {author} {\bibfnamefont {R.}~\bibnamefont {Sekine}}, \bibinfo {author} {\bibfnamefont {Q.}~\bibnamefont {Guo}}, \bibinfo {author} {\bibfnamefont {R.~M.}\ \bibnamefont {Briggs}}, \bibinfo {author} {\bibfnamefont {A.}~\bibnamefont {Marandi}},\ and\ \bibinfo {author} {\bibfnamefont {S.~A.}\ \bibnamefont {Diddams}},\ }\bibfield  {title} {\enquote {\bibinfo {title} {Visible-to-ultraviolet frequency comb generation in lithium niobate nanophotonic waveguides},}\ }\href@noop {} {\bibfield  {journal} {\bibinfo  {journal} {Nature Photonics}\ }\textbf {\bibinfo {volume} {18}},\ \bibinfo {pages} {218--223} (\bibinfo {year} {2024})}\BibitemShut {NoStop}%
\bibitem [{\citenamefont {Tang}\ \emph {et~al.}(2025{\natexlab{a}})\citenamefont {Tang}, \citenamefont {Qiu}, \citenamefont {Ding}, \citenamefont {Ding}, \citenamefont {Song}, \citenamefont {Xian}, \citenamefont {Li}, \citenamefont {Liu}, \citenamefont {Yuan}, \citenamefont {Zheng},\ and\ \citenamefont {Chen}}]{Tang:25}%
  \BibitemOpen
  \bibfield  {author} {\bibinfo {author} {\bibfnamefont {Y.}~\bibnamefont {Tang}}, \bibinfo {author} {\bibfnamefont {J.}~\bibnamefont {Qiu}}, \bibinfo {author} {\bibfnamefont {W.}~\bibnamefont {Ding}}, \bibinfo {author} {\bibfnamefont {T.}~\bibnamefont {Ding}}, \bibinfo {author} {\bibfnamefont {X.}~\bibnamefont {Song}}, \bibinfo {author} {\bibfnamefont {T.}~\bibnamefont {Xian}}, \bibinfo {author} {\bibfnamefont {H.}~\bibnamefont {Li}}, \bibinfo {author} {\bibfnamefont {S.}~\bibnamefont {Liu}}, \bibinfo {author} {\bibfnamefont {L.}~\bibnamefont {Yuan}}, \bibinfo {author} {\bibfnamefont {Y.}~\bibnamefont {Zheng}},\ and\ \bibinfo {author} {\bibfnamefont {X.}~\bibnamefont {Chen}},\ }\bibfield  {title} {\enquote {\bibinfo {title} {Lithium niobate micro-waveguides for efficient supercontinuum generation and frequency comb self-referencing},}\ }\href@noop {} {\bibfield  {journal} {\bibinfo  {journal} {Photon. Res.}\ }\textbf {\bibinfo {volume} {13}},\ \bibinfo {pages} {3332--3340} (\bibinfo {year}
  {2025}{\natexlab{a}})}\BibitemShut {NoStop}%
\bibitem [{\citenamefont {Reig~Escalé}\ \emph {et~al.}(2020)\citenamefont {Reig~Escalé}, \citenamefont {Kaufmann}, \citenamefont {Jiang}, \citenamefont {Pohl},\ and\ \citenamefont {Grange}}]{10.1063/5.0028776}%
  \BibitemOpen
  \bibfield  {author} {\bibinfo {author} {\bibfnamefont {M.}~\bibnamefont {Reig~Escalé}}, \bibinfo {author} {\bibfnamefont {F.}~\bibnamefont {Kaufmann}}, \bibinfo {author} {\bibfnamefont {H.}~\bibnamefont {Jiang}}, \bibinfo {author} {\bibfnamefont {D.}~\bibnamefont {Pohl}},\ and\ \bibinfo {author} {\bibfnamefont {R.}~\bibnamefont {Grange}},\ }\bibfield  {title} {\enquote {\bibinfo {title} {Generation of 280 thz-spanning near-ultraviolet light in lithium niobate-on-insulator waveguides with sub-100 pj pulses},}\ }\href@noop {} {\bibfield  {journal} {\bibinfo  {journal} {APL Photonics}\ }\textbf {\bibinfo {volume} {5}},\ \bibinfo {pages} {121301} (\bibinfo {year} {2020})}\BibitemShut {NoStop}%
\bibitem [{\citenamefont {Ayhan}\ \emph {et~al.}(2025)\citenamefont {Ayhan}, \citenamefont {Ludwig}, \citenamefont {Herr}, \citenamefont {Brasch},\ and\ \citenamefont {Villanueva}}]{ayhan2025fabrication}%
  \BibitemOpen
  \bibfield  {author} {\bibinfo {author} {\bibfnamefont {F.}~\bibnamefont {Ayhan}}, \bibinfo {author} {\bibfnamefont {M.}~\bibnamefont {Ludwig}}, \bibinfo {author} {\bibfnamefont {T.}~\bibnamefont {Herr}}, \bibinfo {author} {\bibfnamefont {V.}~\bibnamefont {Brasch}},\ and\ \bibinfo {author} {\bibfnamefont {L.~G.}\ \bibnamefont {Villanueva}},\ }\bibfield  {title} {\enquote {\bibinfo {title} {Fabrication of periodically poled lithium niobate waveguides for broadband nonlinear photonics},}\ }\href@noop {} {\bibfield  {journal} {\bibinfo  {journal} {APL Photonics}\ }\textbf {\bibinfo {volume} {10}} (\bibinfo {year} {2025})}\BibitemShut {NoStop}%
\bibitem [{\citenamefont {Ludwig}\ \emph {et~al.}(2025)\citenamefont {Ludwig}, \citenamefont {Ayhan}, \citenamefont {Voumard}, \citenamefont {Fan}, \citenamefont {Gaafar}, \citenamefont {Brasch}, \citenamefont {Villanueva},\ and\ \citenamefont {Herr}}]{ludwig2025mid}%
  \BibitemOpen
  \bibfield  {author} {\bibinfo {author} {\bibfnamefont {M.}~\bibnamefont {Ludwig}}, \bibinfo {author} {\bibfnamefont {F.}~\bibnamefont {Ayhan}}, \bibinfo {author} {\bibfnamefont {T.}~\bibnamefont {Voumard}}, \bibinfo {author} {\bibfnamefont {W.}~\bibnamefont {Fan}}, \bibinfo {author} {\bibfnamefont {M.~A.}\ \bibnamefont {Gaafar}}, \bibinfo {author} {\bibfnamefont {V.}~\bibnamefont {Brasch}}, \bibinfo {author} {\bibfnamefont {L.~G.}\ \bibnamefont {Villanueva}},\ and\ \bibinfo {author} {\bibfnamefont {T.}~\bibnamefont {Herr}},\ }\bibfield  {title} {\enquote {\bibinfo {title} {Mid-infrared continua via spectral broadening and difference frequency generation in a nanophotonic lithium niobate waveguide},}\ }\href@noop {} {\bibfield  {journal} {\bibinfo  {journal} {arXiv preprint arXiv:2510.23878}\ } (\bibinfo {year} {2025})}\BibitemShut {NoStop}%
\bibitem [{\citenamefont {Tang}\ \emph {et~al.}(2025{\natexlab{b}})\citenamefont {Tang}, \citenamefont {Yu}, \citenamefont {Zhou}, \citenamefont {Zhang}, \citenamefont {Lu}, \citenamefont {Chen}, \citenamefont {Zhu},\ and\ \citenamefont {Wang}}]{tang2025chip}%
  \BibitemOpen
  \bibfield  {author} {\bibinfo {author} {\bibfnamefont {T.}~\bibnamefont {Tang}}, \bibinfo {author} {\bibfnamefont {S.}~\bibnamefont {Yu}}, \bibinfo {author} {\bibfnamefont {R.}~\bibnamefont {Zhou}}, \bibinfo {author} {\bibfnamefont {J.}~\bibnamefont {Zhang}}, \bibinfo {author} {\bibfnamefont {J.}~\bibnamefont {Lu}}, \bibinfo {author} {\bibfnamefont {G.}~\bibnamefont {Chen}}, \bibinfo {author} {\bibfnamefont {T.}~\bibnamefont {Zhu}},\ and\ \bibinfo {author} {\bibfnamefont {L.}~\bibnamefont {Wang}},\ }\bibfield  {title} {\enquote {\bibinfo {title} {On-chip amplification-free fceo detection and broadband scg in parabolically width-modulated tfln waveguides},}\ }\href@noop {} {\bibfield  {journal} {\bibinfo  {journal} {arXiv preprint arXiv:2509.00394}\ } (\bibinfo {year} {2025}{\natexlab{b}})}\BibitemShut {NoStop}%
\bibitem [{\citenamefont {Fang}, \citenamefont {Shentu},\ and\ \citenamefont {Lu}(2026)}]{fang2026broadband1}%
  \BibitemOpen
  \bibfield  {author} {\bibinfo {author} {\bibfnamefont {X.-X.}\ \bibnamefont {Fang}}, \bibinfo {author} {\bibfnamefont {G.}~\bibnamefont {Shentu}},\ and\ \bibinfo {author} {\bibfnamefont {H.}~\bibnamefont {Lu}},\ }\bibfield  {title} {\enquote {\bibinfo {title} {Broadband quantum photon source in a step-chirped periodically poled lithium niobate waveguide},}\ }\href@noop {} {\bibfield  {journal} {\bibinfo  {journal} {Optics Express}\ }\textbf {\bibinfo {volume} {34}},\ \bibinfo {pages} {3759--3767} (\bibinfo {year} {2026})}\BibitemShut {NoStop}%
\bibitem [{\citenamefont {Gao}\ \emph {et~al.}(2025)\citenamefont {Gao}, \citenamefont {Sun}, \citenamefont {Rebolledo-Salgado}, \citenamefont {Van~Laer}, \citenamefont {Torres-Company},\ and\ \citenamefont {Schr{\"o}der}}]{gao2025tightly}%
  \BibitemOpen
  \bibfield  {author} {\bibinfo {author} {\bibfnamefont {Y.}~\bibnamefont {Gao}}, \bibinfo {author} {\bibfnamefont {Y.}~\bibnamefont {Sun}}, \bibinfo {author} {\bibfnamefont {I.}~\bibnamefont {Rebolledo-Salgado}}, \bibinfo {author} {\bibfnamefont {R.}~\bibnamefont {Van~Laer}}, \bibinfo {author} {\bibfnamefont {V.}~\bibnamefont {Torres-Company}},\ and\ \bibinfo {author} {\bibfnamefont {J.}~\bibnamefont {Schr{\"o}der}},\ }\bibfield  {title} {\enquote {\bibinfo {title} {Tightly-confined and long z-cut lithium niobate waveguide with ultralow-loss},}\ }\href@noop {} {\bibfield  {journal} {\bibinfo  {journal} {Laser \& Photonics Reviews}\ }\textbf {\bibinfo {volume} {19}},\ \bibinfo {pages} {e00042} (\bibinfo {year} {2025})}\BibitemShut {NoStop}%
\bibitem [{\citenamefont {Shi}\ \emph {et~al.}(2025{\natexlab{a}})\citenamefont {Shi}, \citenamefont {Li}, \citenamefont {Du}, \citenamefont {Zhou}, \citenamefont {Yang}, \citenamefont {Lim}, \citenamefont {Mohanraj}, \citenamefont {Zhao}, \citenamefont {Chen}, \citenamefont {Wang} \emph {et~al.}}]{shi2025integrated}%
  \BibitemOpen
  \bibfield  {author} {\bibinfo {author} {\bibfnamefont {X.}~\bibnamefont {Shi}}, \bibinfo {author} {\bibfnamefont {Y.}~\bibnamefont {Li}}, \bibinfo {author} {\bibfnamefont {J.}~\bibnamefont {Du}}, \bibinfo {author} {\bibfnamefont {L.}~\bibnamefont {Zhou}}, \bibinfo {author} {\bibfnamefont {R.}~\bibnamefont {Yang}}, \bibinfo {author} {\bibfnamefont {E.~T.}\ \bibnamefont {Lim}}, \bibinfo {author} {\bibfnamefont {S.~S.}\ \bibnamefont {Mohanraj}}, \bibinfo {author} {\bibfnamefont {M.}~\bibnamefont {Zhao}}, \bibinfo {author} {\bibfnamefont {X.}~\bibnamefont {Chen}}, \bibinfo {author} {\bibfnamefont {X.}~\bibnamefont {Wang}}, \emph {et~al.},\ }\bibfield  {title} {\enquote {\bibinfo {title} {Integrated polarization-entangled photon source for wavelength-multiplexed quantum networks},}\ }\href@noop {} {\bibfield  {journal} {\bibinfo  {journal} {arXiv preprint arXiv:2511.22680}\ } (\bibinfo {year} {2025}{\natexlab{a}})}\BibitemShut {NoStop}%
\bibitem [{\citenamefont {Bollmers}\ \emph {et~al.}(2025)\citenamefont {Bollmers}, \citenamefont {Spiegelberg}, \citenamefont {R{\"u}sing}, \citenamefont {Eigner}, \citenamefont {Padberg},\ and\ \citenamefont {Silberhorn}}]{bollmers2025segmented}%
  \BibitemOpen
  \bibfield  {author} {\bibinfo {author} {\bibfnamefont {L.}~\bibnamefont {Bollmers}}, \bibinfo {author} {\bibfnamefont {N.}~\bibnamefont {Spiegelberg}}, \bibinfo {author} {\bibfnamefont {M.}~\bibnamefont {R{\"u}sing}}, \bibinfo {author} {\bibfnamefont {C.}~\bibnamefont {Eigner}}, \bibinfo {author} {\bibfnamefont {L.}~\bibnamefont {Padberg}},\ and\ \bibinfo {author} {\bibfnamefont {C.}~\bibnamefont {Silberhorn}},\ }\bibfield  {title} {\enquote {\bibinfo {title} {Segmented finger electrodes to optimize ultra-long continuous wafer-scale periodic poling in thin-film lithium niobate},}\ }\href@noop {} {\bibfield  {journal} {\bibinfo  {journal} {Nanophotonics}\ }\textbf {\bibinfo {volume} {14}},\ \bibinfo {pages} {4761--4771} (\bibinfo {year} {2025})}\BibitemShut {NoStop}%
\bibitem [{\citenamefont {Boyd}, \citenamefont {Gaeta},\ and\ \citenamefont {Giese}(2008)}]{boyd2008nonlinear}%
  \BibitemOpen
  \bibfield  {author} {\bibinfo {author} {\bibfnamefont {R.~W.}\ \bibnamefont {Boyd}}, \bibinfo {author} {\bibfnamefont {A.~L.}\ \bibnamefont {Gaeta}},\ and\ \bibinfo {author} {\bibfnamefont {E.}~\bibnamefont {Giese}},\ }\bibfield  {title} {\enquote {\bibinfo {title} {Nonlinear optics},}\ }in\ \href@noop {} {\emph {\bibinfo {booktitle} {Springer Handbook of Atomic, Molecular, and Optical Physics}}}\ (\bibinfo  {publisher} {Springer},\ \bibinfo {year} {2008})\ pp.\ \bibinfo {pages} {1097--1110}\BibitemShut {NoStop}%
\bibitem [{\citenamefont {Fejer}\ \emph {et~al.}(1992)\citenamefont {Fejer}, \citenamefont {Magel}, \citenamefont {Jundt},\ and\ \citenamefont {Byer}}]{fejer1992quasi}%
  \BibitemOpen
  \bibfield  {author} {\bibinfo {author} {\bibfnamefont {M.~M.}\ \bibnamefont {Fejer}}, \bibinfo {author} {\bibfnamefont {G.}~\bibnamefont {Magel}}, \bibinfo {author} {\bibfnamefont {D.~H.}\ \bibnamefont {Jundt}},\ and\ \bibinfo {author} {\bibfnamefont {R.~L.}\ \bibnamefont {Byer}},\ }\bibfield  {title} {\enquote {\bibinfo {title} {Quasi-phase-matched second harmonic generation: tuning and tolerances},}\ }\href@noop {} {\bibfield  {journal} {\bibinfo  {journal} {IEEE Journal of quantum electronics}\ }\textbf {\bibinfo {volume} {28}},\ \bibinfo {pages} {2631--2654} (\bibinfo {year} {1992})}\BibitemShut {NoStop}%
\bibitem [{\citenamefont {Leo}\ \emph {et~al.}(2014)\citenamefont {Leo}, \citenamefont {Gorza}, \citenamefont {Safioui}, \citenamefont {Kockaert}, \citenamefont {Coen}, \citenamefont {Dave}, \citenamefont {Kuyken},\ and\ \citenamefont {Roelkens}}]{leo2014dispersive}%
  \BibitemOpen
  \bibfield  {author} {\bibinfo {author} {\bibfnamefont {F.}~\bibnamefont {Leo}}, \bibinfo {author} {\bibfnamefont {S.-P.}\ \bibnamefont {Gorza}}, \bibinfo {author} {\bibfnamefont {J.}~\bibnamefont {Safioui}}, \bibinfo {author} {\bibfnamefont {P.}~\bibnamefont {Kockaert}}, \bibinfo {author} {\bibfnamefont {S.}~\bibnamefont {Coen}}, \bibinfo {author} {\bibfnamefont {U.}~\bibnamefont {Dave}}, \bibinfo {author} {\bibfnamefont {B.}~\bibnamefont {Kuyken}},\ and\ \bibinfo {author} {\bibfnamefont {G.}~\bibnamefont {Roelkens}},\ }\bibfield  {title} {\enquote {\bibinfo {title} {Dispersive wave emission and supercontinuum generation in a silicon wire waveguide pumped around the 1550 nm telecommunication wavelength},}\ }\href@noop {} {\bibfield  {journal} {\bibinfo  {journal} {Optics letters}\ }\textbf {\bibinfo {volume} {39}},\ \bibinfo {pages} {3623--3626} (\bibinfo {year} {2014})}\BibitemShut {NoStop}%
\bibitem [{\citenamefont {Singh}\ \emph {et~al.}(2018)\citenamefont {Singh}, \citenamefont {Xin}, \citenamefont {Vermeulen}, \citenamefont {Shtyrkova}, \citenamefont {Li}, \citenamefont {Callahan}, \citenamefont {Magden}, \citenamefont {Ruocco}, \citenamefont {Fahrenkopf}, \citenamefont {Baiocco} \emph {et~al.}}]{singh2018octave}%
  \BibitemOpen
  \bibfield  {author} {\bibinfo {author} {\bibfnamefont {N.}~\bibnamefont {Singh}}, \bibinfo {author} {\bibfnamefont {M.}~\bibnamefont {Xin}}, \bibinfo {author} {\bibfnamefont {D.}~\bibnamefont {Vermeulen}}, \bibinfo {author} {\bibfnamefont {K.}~\bibnamefont {Shtyrkova}}, \bibinfo {author} {\bibfnamefont {N.}~\bibnamefont {Li}}, \bibinfo {author} {\bibfnamefont {P.~T.}\ \bibnamefont {Callahan}}, \bibinfo {author} {\bibfnamefont {E.~S.}\ \bibnamefont {Magden}}, \bibinfo {author} {\bibfnamefont {A.}~\bibnamefont {Ruocco}}, \bibinfo {author} {\bibfnamefont {N.}~\bibnamefont {Fahrenkopf}}, \bibinfo {author} {\bibfnamefont {C.}~\bibnamefont {Baiocco}}, \emph {et~al.},\ }\bibfield  {title} {\enquote {\bibinfo {title} {Octave-spanning coherent supercontinuum generation in silicon on insulator from 1.06 $\mu$m to beyond 2.4 $\mu$m},}\ }\href@noop {} {\bibfield  {journal} {\bibinfo  {journal} {Light: Science \& Applications}\ }\textbf {\bibinfo {volume} {7}},\ \bibinfo {pages} {17131--17131} (\bibinfo {year}
  {2018})}\BibitemShut {NoStop}%
\bibitem [{\citenamefont {Yoon~Oh}\ \emph {et~al.}(2017)\citenamefont {Yoon~Oh}, \citenamefont {Yang}, \citenamefont {Fredrick}, \citenamefont {Ycas}, \citenamefont {Diddams},\ and\ \citenamefont {Vahala}}]{yoon2017coherent}%
  \BibitemOpen
  \bibfield  {author} {\bibinfo {author} {\bibfnamefont {D.}~\bibnamefont {Yoon~Oh}}, \bibinfo {author} {\bibfnamefont {K.~Y.}\ \bibnamefont {Yang}}, \bibinfo {author} {\bibfnamefont {C.}~\bibnamefont {Fredrick}}, \bibinfo {author} {\bibfnamefont {G.}~\bibnamefont {Ycas}}, \bibinfo {author} {\bibfnamefont {S.~A.}\ \bibnamefont {Diddams}},\ and\ \bibinfo {author} {\bibfnamefont {K.~J.}\ \bibnamefont {Vahala}},\ }\bibfield  {title} {\enquote {\bibinfo {title} {Coherent ultra-violet to near-infrared generation in silica ridge waveguides},}\ }\href@noop {} {\bibfield  {journal} {\bibinfo  {journal} {Nature communications}\ }\textbf {\bibinfo {volume} {8}},\ \bibinfo {pages} {13922} (\bibinfo {year} {2017})}\BibitemShut {NoStop}%
\bibitem [{\citenamefont {Grassani}\ \emph {et~al.}(2019)\citenamefont {Grassani}, \citenamefont {Tagkoudi}, \citenamefont {Guo}, \citenamefont {Herkommer}, \citenamefont {Yang}, \citenamefont {Kippenberg},\ and\ \citenamefont {Br{\`e}s}}]{grassani2019mid}%
  \BibitemOpen
  \bibfield  {author} {\bibinfo {author} {\bibfnamefont {D.}~\bibnamefont {Grassani}}, \bibinfo {author} {\bibfnamefont {E.}~\bibnamefont {Tagkoudi}}, \bibinfo {author} {\bibfnamefont {H.}~\bibnamefont {Guo}}, \bibinfo {author} {\bibfnamefont {C.}~\bibnamefont {Herkommer}}, \bibinfo {author} {\bibfnamefont {F.}~\bibnamefont {Yang}}, \bibinfo {author} {\bibfnamefont {T.~J.}\ \bibnamefont {Kippenberg}},\ and\ \bibinfo {author} {\bibfnamefont {C.-S.}\ \bibnamefont {Br{\`e}s}},\ }\bibfield  {title} {\enquote {\bibinfo {title} {Mid infrared gas spectroscopy using efficient fiber laser driven photonic chip-based supercontinuum},}\ }\href@noop {} {\bibfield  {journal} {\bibinfo  {journal} {Nature communications}\ }\textbf {\bibinfo {volume} {10}},\ \bibinfo {pages} {1553} (\bibinfo {year} {2019})}\BibitemShut {NoStop}%
\bibitem [{\citenamefont {Porcel}\ \emph {et~al.}(2017)\citenamefont {Porcel}, \citenamefont {Schepers}, \citenamefont {Epping}, \citenamefont {Hellwig}, \citenamefont {Hoekman}, \citenamefont {Heideman}, \citenamefont {van~der Slot}, \citenamefont {Lee}, \citenamefont {Schmidt}, \citenamefont {Bratschitsch} \emph {et~al.}}]{porcel2017two}%
  \BibitemOpen
  \bibfield  {author} {\bibinfo {author} {\bibfnamefont {M.~A.}\ \bibnamefont {Porcel}}, \bibinfo {author} {\bibfnamefont {F.}~\bibnamefont {Schepers}}, \bibinfo {author} {\bibfnamefont {J.~P.}\ \bibnamefont {Epping}}, \bibinfo {author} {\bibfnamefont {T.}~\bibnamefont {Hellwig}}, \bibinfo {author} {\bibfnamefont {M.}~\bibnamefont {Hoekman}}, \bibinfo {author} {\bibfnamefont {R.~G.}\ \bibnamefont {Heideman}}, \bibinfo {author} {\bibfnamefont {P.~J.}\ \bibnamefont {van~der Slot}}, \bibinfo {author} {\bibfnamefont {C.~J.}\ \bibnamefont {Lee}}, \bibinfo {author} {\bibfnamefont {R.}~\bibnamefont {Schmidt}}, \bibinfo {author} {\bibfnamefont {R.}~\bibnamefont {Bratschitsch}}, \emph {et~al.},\ }\bibfield  {title} {\enquote {\bibinfo {title} {Two-octave spanning supercontinuum generation in stoichiometric silicon nitride waveguides pumped at telecom wavelengths},}\ }\href@noop {} {\bibfield  {journal} {\bibinfo  {journal} {Optics express}\ }\textbf {\bibinfo {volume} {25}},\ \bibinfo {pages} {1542--1554} (\bibinfo
  {year} {2017})}\BibitemShut {NoStop}%
\bibitem [{\citenamefont {Deniel}\ \emph {et~al.}(2023)\citenamefont {Deniel}, \citenamefont {Guidry}, \citenamefont {Lukin}, \citenamefont {Yang}, \citenamefont {Yang}, \citenamefont {Vu{\v{c}}kovi{\'c}}, \citenamefont {H{\"a}nsch},\ and\ \citenamefont {Picqu{\'e}}}]{deniel2023visible}%
  \BibitemOpen
  \bibfield  {author} {\bibinfo {author} {\bibfnamefont {L.}~\bibnamefont {Deniel}}, \bibinfo {author} {\bibfnamefont {M.~A.}\ \bibnamefont {Guidry}}, \bibinfo {author} {\bibfnamefont {D.~M.}\ \bibnamefont {Lukin}}, \bibinfo {author} {\bibfnamefont {K.~Y.}\ \bibnamefont {Yang}}, \bibinfo {author} {\bibfnamefont {J.}~\bibnamefont {Yang}}, \bibinfo {author} {\bibfnamefont {J.}~\bibnamefont {Vu{\v{c}}kovi{\'c}}}, \bibinfo {author} {\bibfnamefont {T.~W.}\ \bibnamefont {H{\"a}nsch}},\ and\ \bibinfo {author} {\bibfnamefont {N.}~\bibnamefont {Picqu{\'e}}},\ }\bibfield  {title} {\enquote {\bibinfo {title} {Visible to mid-infrared supercontinuum generation in 4h-silicon-carbide nanophotonic waveguides},}\ }in\ \href@noop {} {\emph {\bibinfo {booktitle} {CLEO: Science and Innovations}}}\ (\bibinfo {organization} {Optica Publishing Group},\ \bibinfo {year} {2023})\ pp.\ \bibinfo {pages} {STh1F--4}\BibitemShut {NoStop}%
\bibitem [{\citenamefont {Kuyken}\ \emph {et~al.}(2020)\citenamefont {Kuyken}, \citenamefont {Billet}, \citenamefont {Leo}, \citenamefont {Yvind},\ and\ \citenamefont {Pu}}]{kuyken2020octave}%
  \BibitemOpen
  \bibfield  {author} {\bibinfo {author} {\bibfnamefont {B.}~\bibnamefont {Kuyken}}, \bibinfo {author} {\bibfnamefont {M.}~\bibnamefont {Billet}}, \bibinfo {author} {\bibfnamefont {F.}~\bibnamefont {Leo}}, \bibinfo {author} {\bibfnamefont {K.}~\bibnamefont {Yvind}},\ and\ \bibinfo {author} {\bibfnamefont {M.}~\bibnamefont {Pu}},\ }\bibfield  {title} {\enquote {\bibinfo {title} {Octave-spanning coherent supercontinuum generation in an algaas-on-insulator waveguide},}\ }\href@noop {} {\bibfield  {journal} {\bibinfo  {journal} {Optics Letters}\ }\textbf {\bibinfo {volume} {45}},\ \bibinfo {pages} {603--606} (\bibinfo {year} {2020})}\BibitemShut {NoStop}%
\bibitem [{\citenamefont {Fan}\ \emph {et~al.}(2024)\citenamefont {Fan}, \citenamefont {Ludwig}, \citenamefont {Rousseau}, \citenamefont {Arabadzhiev}, \citenamefont {Ruhnke}, \citenamefont {Wildi},\ and\ \citenamefont {Herr}}]{fan2024supercontinua}%
  \BibitemOpen
  \bibfield  {author} {\bibinfo {author} {\bibfnamefont {W.}~\bibnamefont {Fan}}, \bibinfo {author} {\bibfnamefont {M.}~\bibnamefont {Ludwig}}, \bibinfo {author} {\bibfnamefont {I.}~\bibnamefont {Rousseau}}, \bibinfo {author} {\bibfnamefont {I.}~\bibnamefont {Arabadzhiev}}, \bibinfo {author} {\bibfnamefont {B.}~\bibnamefont {Ruhnke}}, \bibinfo {author} {\bibfnamefont {T.}~\bibnamefont {Wildi}},\ and\ \bibinfo {author} {\bibfnamefont {T.}~\bibnamefont {Herr}},\ }\bibfield  {title} {\enquote {\bibinfo {title} {Supercontinua from integrated gallium nitride waveguides},}\ }\href@noop {} {\bibfield  {journal} {\bibinfo  {journal} {Optica}\ }\textbf {\bibinfo {volume} {11}},\ \bibinfo {pages} {1175--1181} (\bibinfo {year} {2024})}\BibitemShut {NoStop}%
\bibitem [{\citenamefont {Dave}\ \emph {et~al.}(2015)\citenamefont {Dave}, \citenamefont {Ciret}, \citenamefont {Gorza}, \citenamefont {Combrie}, \citenamefont {De~Rossi}, \citenamefont {Raineri}, \citenamefont {Roelkens},\ and\ \citenamefont {Kuyken}}]{dave2015dispersive}%
  \BibitemOpen
  \bibfield  {author} {\bibinfo {author} {\bibfnamefont {U.~D.}\ \bibnamefont {Dave}}, \bibinfo {author} {\bibfnamefont {C.}~\bibnamefont {Ciret}}, \bibinfo {author} {\bibfnamefont {S.-P.}\ \bibnamefont {Gorza}}, \bibinfo {author} {\bibfnamefont {S.}~\bibnamefont {Combrie}}, \bibinfo {author} {\bibfnamefont {A.}~\bibnamefont {De~Rossi}}, \bibinfo {author} {\bibfnamefont {F.}~\bibnamefont {Raineri}}, \bibinfo {author} {\bibfnamefont {G.}~\bibnamefont {Roelkens}},\ and\ \bibinfo {author} {\bibfnamefont {B.}~\bibnamefont {Kuyken}},\ }\bibfield  {title} {\enquote {\bibinfo {title} {Dispersive-wave-based octave-spanning supercontinuum generation in ingap membrane waveguides on a silicon substrate},}\ }\href@noop {} {\bibfield  {journal} {\bibinfo  {journal} {Optics letters}\ }\textbf {\bibinfo {volume} {40}},\ \bibinfo {pages} {3584--3587} (\bibinfo {year} {2015})}\BibitemShut {NoStop}%
\bibitem [{\citenamefont {Jung}\ \emph {et~al.}(2021)\citenamefont {Jung}, \citenamefont {Yu}, \citenamefont {Carlson}, \citenamefont {Drake}, \citenamefont {Briles},\ and\ \citenamefont {Papp}}]{jung2021tantala}%
  \BibitemOpen
  \bibfield  {author} {\bibinfo {author} {\bibfnamefont {H.}~\bibnamefont {Jung}}, \bibinfo {author} {\bibfnamefont {S.-P.}\ \bibnamefont {Yu}}, \bibinfo {author} {\bibfnamefont {D.~R.}\ \bibnamefont {Carlson}}, \bibinfo {author} {\bibfnamefont {T.~E.}\ \bibnamefont {Drake}}, \bibinfo {author} {\bibfnamefont {T.~C.}\ \bibnamefont {Briles}},\ and\ \bibinfo {author} {\bibfnamefont {S.~B.}\ \bibnamefont {Papp}},\ }\bibfield  {title} {\enquote {\bibinfo {title} {Tantala kerr nonlinear integrated photonics},}\ }\href@noop {} {\bibfield  {journal} {\bibinfo  {journal} {Optica}\ }\textbf {\bibinfo {volume} {8}},\ \bibinfo {pages} {811--817} (\bibinfo {year} {2021})}\BibitemShut {NoStop}%
\bibitem [{\citenamefont {Singh}\ \emph {et~al.}(2020)\citenamefont {Singh}, \citenamefont {Mbonde}, \citenamefont {Frankis}, \citenamefont {Ippen}, \citenamefont {Bradley},\ and\ \citenamefont {K{\"a}rtner}}]{singh2020nonlinear}%
  \BibitemOpen
  \bibfield  {author} {\bibinfo {author} {\bibfnamefont {N.}~\bibnamefont {Singh}}, \bibinfo {author} {\bibfnamefont {H.~M.}\ \bibnamefont {Mbonde}}, \bibinfo {author} {\bibfnamefont {H.~C.}\ \bibnamefont {Frankis}}, \bibinfo {author} {\bibfnamefont {E.}~\bibnamefont {Ippen}}, \bibinfo {author} {\bibfnamefont {J.~D.}\ \bibnamefont {Bradley}},\ and\ \bibinfo {author} {\bibfnamefont {F.~X.}\ \bibnamefont {K{\"a}rtner}},\ }\bibfield  {title} {\enquote {\bibinfo {title} {Nonlinear silicon photonics on cmos-compatible tellurium oxide},}\ }\href@noop {} {\bibfield  {journal} {\bibinfo  {journal} {Photonics Research}\ }\textbf {\bibinfo {volume} {8}},\ \bibinfo {pages} {1904--1909} (\bibinfo {year} {2020})}\BibitemShut {NoStop}%
\bibitem [{\citenamefont {Hammani}\ \emph {et~al.}(2018)\citenamefont {Hammani}, \citenamefont {Markey}, \citenamefont {Lamy}, \citenamefont {Kibler}, \citenamefont {Arocas}, \citenamefont {Fatome}, \citenamefont {Dereux}, \citenamefont {Weeber},\ and\ \citenamefont {Finot}}]{hammani2018octave}%
  \BibitemOpen
  \bibfield  {author} {\bibinfo {author} {\bibfnamefont {K.}~\bibnamefont {Hammani}}, \bibinfo {author} {\bibfnamefont {L.}~\bibnamefont {Markey}}, \bibinfo {author} {\bibfnamefont {M.}~\bibnamefont {Lamy}}, \bibinfo {author} {\bibfnamefont {B.}~\bibnamefont {Kibler}}, \bibinfo {author} {\bibfnamefont {J.}~\bibnamefont {Arocas}}, \bibinfo {author} {\bibfnamefont {J.}~\bibnamefont {Fatome}}, \bibinfo {author} {\bibfnamefont {A.}~\bibnamefont {Dereux}}, \bibinfo {author} {\bibfnamefont {J.-C.}\ \bibnamefont {Weeber}},\ and\ \bibinfo {author} {\bibfnamefont {C.}~\bibnamefont {Finot}},\ }\bibfield  {title} {\enquote {\bibinfo {title} {Octave spanning supercontinuum in titanium dioxide waveguides},}\ }\href@noop {} {\bibfield  {journal} {\bibinfo  {journal} {Applied Sciences}\ }\textbf {\bibinfo {volume} {8}},\ \bibinfo {pages} {543} (\bibinfo {year} {2018})}\BibitemShut {NoStop}%
\bibitem [{\citenamefont {Hwang}\ \emph {et~al.}(2021)\citenamefont {Hwang}, \citenamefont {Kim}, \citenamefont {Han}, \citenamefont {Jeong}, \citenamefont {Lee}, \citenamefont {Choi},\ and\ \citenamefont {Lee}}]{hwang2021supercontinuum}%
  \BibitemOpen
  \bibfield  {author} {\bibinfo {author} {\bibfnamefont {J.}~\bibnamefont {Hwang}}, \bibinfo {author} {\bibfnamefont {D.-G.}\ \bibnamefont {Kim}}, \bibinfo {author} {\bibfnamefont {S.}~\bibnamefont {Han}}, \bibinfo {author} {\bibfnamefont {D.}~\bibnamefont {Jeong}}, \bibinfo {author} {\bibfnamefont {Y.-H.}\ \bibnamefont {Lee}}, \bibinfo {author} {\bibfnamefont {D.-Y.}\ \bibnamefont {Choi}},\ and\ \bibinfo {author} {\bibfnamefont {H.}~\bibnamefont {Lee}},\ }\bibfield  {title} {\enquote {\bibinfo {title} {Supercontinuum generation in as2s3 waveguides fabricated without direct etching},}\ }\href@noop {} {\bibfield  {journal} {\bibinfo  {journal} {Optics Letters}\ }\textbf {\bibinfo {volume} {46}},\ \bibinfo {pages} {2413--2416} (\bibinfo {year} {2021})}\BibitemShut {NoStop}%
\bibitem [{\citenamefont {Choi}\ \emph {et~al.}(2016)\citenamefont {Choi}, \citenamefont {Han}, \citenamefont {Sohn}, \citenamefont {Chen}, \citenamefont {Smith}, \citenamefont {Kimerling}, \citenamefont {Richardson}, \citenamefont {Agarwal},\ and\ \citenamefont {Tan}}]{choi2016nonlinear}%
  \BibitemOpen
  \bibfield  {author} {\bibinfo {author} {\bibfnamefont {J.~W.}\ \bibnamefont {Choi}}, \bibinfo {author} {\bibfnamefont {Z.}~\bibnamefont {Han}}, \bibinfo {author} {\bibfnamefont {B.-U.}\ \bibnamefont {Sohn}}, \bibinfo {author} {\bibfnamefont {G.~F.}\ \bibnamefont {Chen}}, \bibinfo {author} {\bibfnamefont {C.}~\bibnamefont {Smith}}, \bibinfo {author} {\bibfnamefont {L.~C.}\ \bibnamefont {Kimerling}}, \bibinfo {author} {\bibfnamefont {K.~A.}\ \bibnamefont {Richardson}}, \bibinfo {author} {\bibfnamefont {A.~M.}\ \bibnamefont {Agarwal}},\ and\ \bibinfo {author} {\bibfnamefont {D.~T.}\ \bibnamefont {Tan}},\ }\bibfield  {title} {\enquote {\bibinfo {title} {Nonlinear characterization of gesbs chalcogenide glass waveguides},}\ }\href@noop {} {\bibfield  {journal} {\bibinfo  {journal} {Scientific Reports}\ }\textbf {\bibinfo {volume} {6}},\ \bibinfo {pages} {39234} (\bibinfo {year} {2016})}\BibitemShut {NoStop}%
\bibitem [{\citenamefont {Wang}\ \emph {et~al.}(2025)\citenamefont {Wang}, \citenamefont {Tang}, \citenamefont {Zhu},\ and\ \citenamefont {Lu}}]{wang2025three}%
  \BibitemOpen
  \bibfield  {author} {\bibinfo {author} {\bibfnamefont {L.}~\bibnamefont {Wang}}, \bibinfo {author} {\bibfnamefont {T.}~\bibnamefont {Tang}}, \bibinfo {author} {\bibfnamefont {H.}~\bibnamefont {Zhu}},\ and\ \bibinfo {author} {\bibfnamefont {J.}~\bibnamefont {Lu}},\ }\bibfield  {title} {\enquote {\bibinfo {title} {Three-octave supercontinuum generation spanning from ultraviolet in lithium tantalate waveguides},}\ }\href@noop {} {\bibfield  {journal} {\bibinfo  {journal} {arXiv preprint arXiv:2512.16350}\ } (\bibinfo {year} {2025})}\BibitemShut {NoStop}%
\bibitem [{\citenamefont {Franken}\ \emph {et~al.}(2025)\citenamefont {Franken}, \citenamefont {Ghosh}, \citenamefont {Rodrigues}, \citenamefont {Yang}, \citenamefont {Xin}, \citenamefont {Lu}, \citenamefont {Witt}, \citenamefont {Joe}, \citenamefont {Wiederhecker}, \citenamefont {Boller} \emph {et~al.}}]{franken2025milliwatt}%
  \BibitemOpen
  \bibfield  {author} {\bibinfo {author} {\bibfnamefont {C.~A.}\ \bibnamefont {Franken}}, \bibinfo {author} {\bibfnamefont {S.~S.}\ \bibnamefont {Ghosh}}, \bibinfo {author} {\bibfnamefont {C.~C.}\ \bibnamefont {Rodrigues}}, \bibinfo {author} {\bibfnamefont {J.}~\bibnamefont {Yang}}, \bibinfo {author} {\bibfnamefont {C.~J.}\ \bibnamefont {Xin}}, \bibinfo {author} {\bibfnamefont {S.}~\bibnamefont {Lu}}, \bibinfo {author} {\bibfnamefont {D.}~\bibnamefont {Witt}}, \bibinfo {author} {\bibfnamefont {G.}~\bibnamefont {Joe}}, \bibinfo {author} {\bibfnamefont {G.}~\bibnamefont {Wiederhecker}}, \bibinfo {author} {\bibfnamefont {K.-J.}\ \bibnamefont {Boller}}, \emph {et~al.},\ }\bibfield  {title} {\enquote {\bibinfo {title} {Milliwatt-level uv generation using sidewall poled lithium niobate},}\ }\href@noop {} {\bibfield  {journal} {\bibinfo  {journal} {arXiv preprint arXiv:2503.16785}\ } (\bibinfo {year} {2025})}\BibitemShut {NoStop}%
\bibitem [{\citenamefont {Shi}\ \emph {et~al.}(2024)\citenamefont {Shi}, \citenamefont {Mohanraj}, \citenamefont {Dhyani}, \citenamefont {Baiju}, \citenamefont {Wang}, \citenamefont {Sun}, \citenamefont {Zhou}, \citenamefont {Paterova}, \citenamefont {Leong},\ and\ \citenamefont {Zhu}}]{shi2024efficient}%
  \BibitemOpen
  \bibfield  {author} {\bibinfo {author} {\bibfnamefont {X.}~\bibnamefont {Shi}}, \bibinfo {author} {\bibfnamefont {S.~S.}\ \bibnamefont {Mohanraj}}, \bibinfo {author} {\bibfnamefont {V.}~\bibnamefont {Dhyani}}, \bibinfo {author} {\bibfnamefont {A.~A.}\ \bibnamefont {Baiju}}, \bibinfo {author} {\bibfnamefont {S.}~\bibnamefont {Wang}}, \bibinfo {author} {\bibfnamefont {J.}~\bibnamefont {Sun}}, \bibinfo {author} {\bibfnamefont {L.}~\bibnamefont {Zhou}}, \bibinfo {author} {\bibfnamefont {A.}~\bibnamefont {Paterova}}, \bibinfo {author} {\bibfnamefont {V.}~\bibnamefont {Leong}},\ and\ \bibinfo {author} {\bibfnamefont {D.}~\bibnamefont {Zhu}},\ }\bibfield  {title} {\enquote {\bibinfo {title} {Efficient photon-pair generation in layer-poled lithium niobate nanophotonic waveguides},}\ }\href@noop {} {\bibfield  {journal} {\bibinfo  {journal} {Light: Science \& Applications}\ }\textbf {\bibinfo {volume} {13}},\ \bibinfo {pages} {282} (\bibinfo {year} {2024})}\BibitemShut {NoStop}%
\bibitem [{\citenamefont {Liang}\ \emph {et~al.}(2022)\citenamefont {Liang}, \citenamefont {Fu}, \citenamefont {Yu}, \citenamefont {Xue}, \citenamefont {Shi}, \citenamefont {Lu}, \citenamefont {Chen}, \citenamefont {Zhang}, \citenamefont {Luo}, \citenamefont {Hu} \emph {et~al.}}]{liang2022efficient}%
  \BibitemOpen
  \bibfield  {author} {\bibinfo {author} {\bibfnamefont {X.}~\bibnamefont {Liang}}, \bibinfo {author} {\bibfnamefont {L.}~\bibnamefont {Fu}}, \bibinfo {author} {\bibfnamefont {Q.}~\bibnamefont {Yu}}, \bibinfo {author} {\bibfnamefont {Z.}~\bibnamefont {Xue}}, \bibinfo {author} {\bibfnamefont {X.}~\bibnamefont {Shi}}, \bibinfo {author} {\bibfnamefont {Y.}~\bibnamefont {Lu}}, \bibinfo {author} {\bibfnamefont {H.}~\bibnamefont {Chen}}, \bibinfo {author} {\bibfnamefont {B.}~\bibnamefont {Zhang}}, \bibinfo {author} {\bibfnamefont {Y.}~\bibnamefont {Luo}}, \bibinfo {author} {\bibfnamefont {Q.}~\bibnamefont {Hu}}, \emph {et~al.},\ }\bibfield  {title} {\enquote {\bibinfo {title} {Efficient and broadband trident spot-size convertor for thin-film lithium niobate integrated device},}\ }\href@noop {} {\bibfield  {journal} {\bibinfo  {journal} {IEEE Photonics Technology Letters}\ }\textbf {\bibinfo {volume} {35}},\ \bibinfo {pages} {35--38} (\bibinfo {year} {2022})}\BibitemShut {NoStop}%
\bibitem [{\citenamefont {Helbing}\ \emph {et~al.}(2002)\citenamefont {Helbing}, \citenamefont {Steinmeyer}, \citenamefont {Stenger}, \citenamefont {Telle},\ and\ \citenamefont {Keller}}]{helbing2002carrier}%
  \BibitemOpen
  \bibfield  {author} {\bibinfo {author} {\bibfnamefont {F.}~\bibnamefont {Helbing}}, \bibinfo {author} {\bibfnamefont {G.}~\bibnamefont {Steinmeyer}}, \bibinfo {author} {\bibfnamefont {J.}~\bibnamefont {Stenger}}, \bibinfo {author} {\bibfnamefont {H.}~\bibnamefont {Telle}},\ and\ \bibinfo {author} {\bibfnamefont {U.}~\bibnamefont {Keller}},\ }\bibfield  {title} {\enquote {\bibinfo {title} {Carrier--envelope-offset dynamics and stabilization of femtosecond pulses},}\ }\href@noop {} {\bibfield  {journal} {\bibinfo  {journal} {Applied Physics B}\ }\textbf {\bibinfo {volume} {74}},\ \bibinfo {pages} {s35--s42} (\bibinfo {year} {2002})}\BibitemShut {NoStop}%
\bibitem [{\citenamefont {Shi}\ \emph {et~al.}(2025{\natexlab{b}})\citenamefont {Shi}, \citenamefont {Baiju}, \citenamefont {Chen}, \citenamefont {Mohanraj}, \citenamefont {Wang}, \citenamefont {Dhyani}, \citenamefont {Shajilal}, \citenamefont {Zhao}, \citenamefont {Yang}, \citenamefont {Li} \emph {et~al.}}]{shi2025squeezed}%
  \BibitemOpen
  \bibfield  {author} {\bibinfo {author} {\bibfnamefont {X.}~\bibnamefont {Shi}}, \bibinfo {author} {\bibfnamefont {A.~A.}\ \bibnamefont {Baiju}}, \bibinfo {author} {\bibfnamefont {X.}~\bibnamefont {Chen}}, \bibinfo {author} {\bibfnamefont {S.~S.}\ \bibnamefont {Mohanraj}}, \bibinfo {author} {\bibfnamefont {S.}~\bibnamefont {Wang}}, \bibinfo {author} {\bibfnamefont {V.}~\bibnamefont {Dhyani}}, \bibinfo {author} {\bibfnamefont {B.}~\bibnamefont {Shajilal}}, \bibinfo {author} {\bibfnamefont {M.}~\bibnamefont {Zhao}}, \bibinfo {author} {\bibfnamefont {R.}~\bibnamefont {Yang}}, \bibinfo {author} {\bibfnamefont {Y.}~\bibnamefont {Li}}, \emph {et~al.},\ }\bibfield  {title} {\enquote {\bibinfo {title} {Squeezed light generation in periodically poled thin-film lithium niobate waveguides},}\ }\href@noop {} {\bibfield  {journal} {\bibinfo  {journal} {Nanophotonics}\ }\textbf {\bibinfo {volume} {14}},\ \bibinfo {pages} {4721--4727} (\bibinfo {year} {2025}{\natexlab{b}})}\BibitemShut {NoStop}%
\end{thebibliography}%

\end{document}


\title{Segment-chirped periodically poled lithium niobate waveguides for broadband supercontinuum generation}

\author{Yue Li} 
\affiliation{Department of Materials Science and Engineering, National University of Singapore, Singapore 117575, Singapore}
\affiliation{College of Electronics and Information Engineering, Sichuan University, Chengdu 610065, China}

\author{Xiaodong Shi}
\affiliation{A$^\ast$STAR Quantum Innovation Centre (Q.InC), Agency for Science, Technology and Research (A$^\ast$STAR), Singapore 138634, Singapore}

\author{Sakthi Sanjeev Mohanraj}
\affiliation{Department of Materials Science and Engineering, National University of Singapore, Singapore 117575, Singapore}

\author{Mengyao Zhao}
\affiliation{Department of Materials Science and Engineering, National University of Singapore, Singapore 117575, Singapore}

\author{Xu Chen}
\affiliation{Department of Materials Science and Engineering, National University of Singapore, Singapore 117575, Singapore}

\author{Xuan Mao}
\affiliation{School of Electrical and Electronic Engineering, Nanyang Technological University, Singapore 639798, Singapore}

\author{Qijie Wang}
\affiliation{School of Electrical and Electronic Engineering, Nanyang Technological University, Singapore 639798, Singapore}

\author{Shouhuan Zhou}
\affiliation{College of Electronics and Information Engineering, Sichuan University, Chengdu 610065, China}

\author{Guoliang Deng}
\affiliation{College of Electronics and Information Engineering, Sichuan University, Chengdu 610065, China}

\author{Di Zhu}
\email{dizhu@nus.edu.sg}
\affiliation{Department of Materials Science and Engineering, National University of Singapore, Singapore 117575, Singapore}
\affiliation{A$^\ast$STAR Quantum Innovation Centre (Q.InC), Agency for Science, Technology and Research (A$^\ast$STAR), Singapore 138634, Singapore}
\affiliation{Centre for Quantum Technologies, National University of Singapore, Singapore 117543, Singapore}
\maketitle





\renewcommand{\thesection}{S\arabic{section}}
\renewcommand{\thefigure}{S\arabic{figure}}

\section{Supercontinuum simulation}

\begin{figure*}[htbp]
\centering 
\includegraphics[width = 6in]{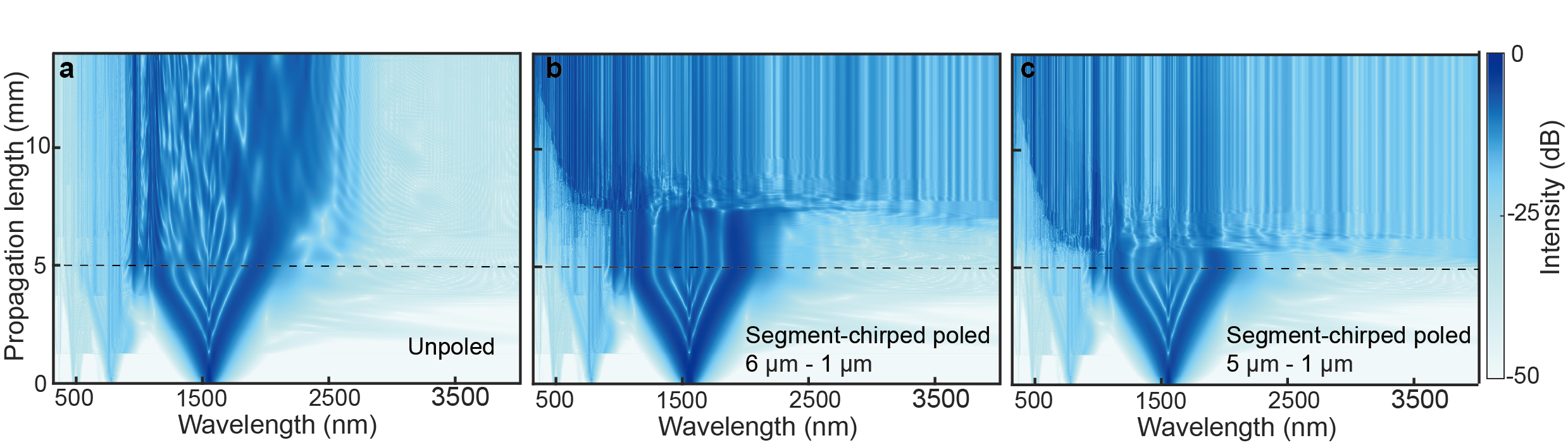}
\caption{Simulation of supercontinuum evolution in \textbf{a,} unpoled TFLN waveguide,  \textbf{b,} SC-PPLN waveguide with periods varying from 6 \textmu m to 1 \textmu m, and \textbf{c,} SC-PPLN waveguide with periods varying from 5 \textmu m to 1 \textmu m, at a pump pulse energy of 477 pJ.}
\label{figs1}
\end{figure*}

To analyze the spectral broadening due to the segmented-chirped periodic poling induced $\chi^{(2)}$ nonlinear interactions, we investigate the spectral evolution in unpoled and poled TFLN waveguides at the same pump pulse energy of 477 pJ \cite{voumard2023simulating}.
In the unpoled waveguide, the spectral broadening is limited in the near-infrared wavelength range, primarily governed by the $\chi^{(3)}$ nonlinear effect, that is self-phase modulation (SPM, Fig.~\ref{figs1}a).
However, in the segment-chirped periodically poled waveguides with poling periods varying from 6 \textmu m to 1 \textmu m, the spectrum broadening covers UV to mid-infrared wavelength range, thanks to the multiple $\chi^{(2)}$ nonlinear interactions (Fig.~\ref{figs1}b).
We then simulate the spectrum in the segment-chirped poled waveguide with poling periods varying from 5 \textmu m to 1 \textmu m, which reveals that the spectrum exhibits a weaker enhancement in the visible and UV wavelength range, due to the lack of third-order quasi-phase matching (QPM) assistance (Fig.~\ref{figs1}c).

\begin{figure*}[htbp]
\centering 
\includegraphics[width = 5.5in]{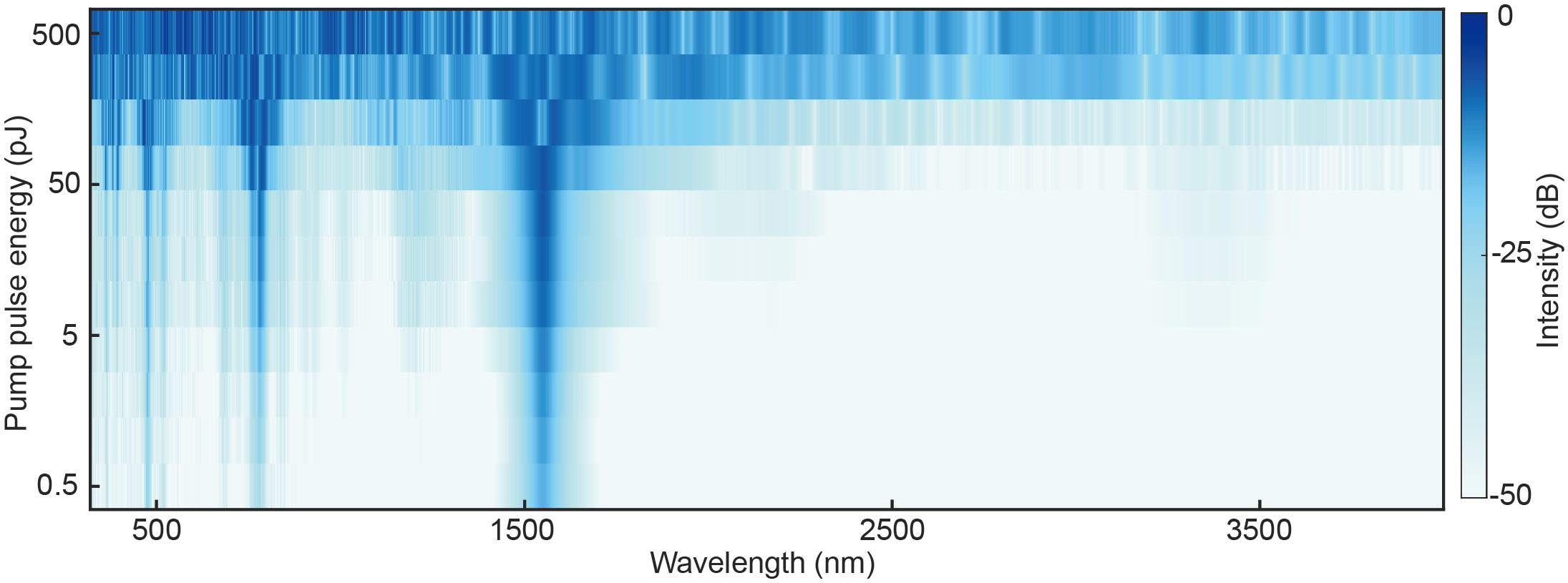}
\caption 
{Simulation of supercontinuum generation from UV to mid-IR with increasing pump pulse energy.}
\label{figs2}
\end{figure*}

We simulate the spectrum broadening with increasing pump pulse energy (Fig. \ref{figs2}).
The results show that when the pump pulse energy exceeds approximately 200~pJ, a broadband supercontinuum covering the entire investigated spectral range is generated.

\section{Wavelength conversion simulation}

The theoretical normalized SHG efficiency is calculated through\cite{wu2022broadband,zhang2022second,fang2026broadband}
\begin{equation}
\eta_{\mathrm{norm}} = \frac{8\pi^{2}}{\varepsilon_{0} c\, n_{\mathrm{FH}}^{2} n_{\mathrm{SH}}\, \lambda_{\omega}^{2}}
\frac{d_{\mathrm{eff}}^{2}}{S_{\mathrm{eff}}}
G^{2}(\Delta k);
\end{equation}
\begin{equation}
S_{\mathrm{eff}}
=
\frac{
\left[ \iint_{all} E_{z,\mathrm{FH}}^{2}(x,z)\, \mathrm{d}x\,\mathrm{d}z \right]^{2}
\iint_{all} E_{z,\mathrm{SH}}^{2}(x,z)\, \mathrm{d}x\,\mathrm{d}z
}{
\left[
\iint_{LN}\,
E_{z,\mathrm{FH}}^{2}(x,z)\,
E_{z,\mathrm{SH}}(x,z)\,
\mathrm{d}x\,\mathrm{d}z
\right]^{2}
};
\end{equation}
\begin{equation}
G^{2}(\Delta k)
=
\left|
\frac{1}{L}
\int_{0}^{L}
d(y)\,
e^{-i\Delta k y}\,
\mathrm{d}y
\right|^{2}.
\end{equation}
Here, $\varepsilon_{0}$ and $c$ are the vacuum permittivity and the speed of light in vacuum, respectively, $n_{\mathrm{FH}}$ and $n_{\mathrm{SH}}$ denote the effective refractive indices of the fundamental-harmonic (FH) and second-harmonic (SH) modes, respectively,
$E_{z,\mathrm{FH}}(x,z)$ and $E_{z,\mathrm{SH}}(x,z)$ represent the horizontal electric-field components of the fundamental TE modes at the FH and SH frequencies, respectively.
The effective nonlinear coefficient is taken as $d_{\mathrm{eff}} = d_{33} = 27~\mathrm{pm/V}$.

\begin{figure}[htbp]
\centering 
\includegraphics[width = 5.5in]{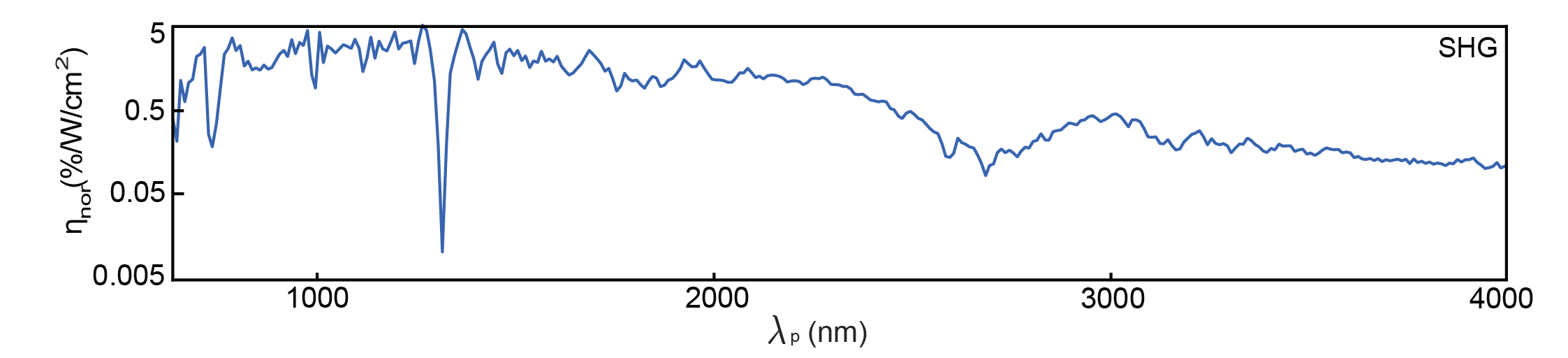}
\caption 
{Simulation of SHG efficiency over a broad wavelength range from 640 nm to 4000 nm.}
\label{figs3}
\end{figure}

\begin{figure*}[htbp]
\centering 
\includegraphics[height=5cm]{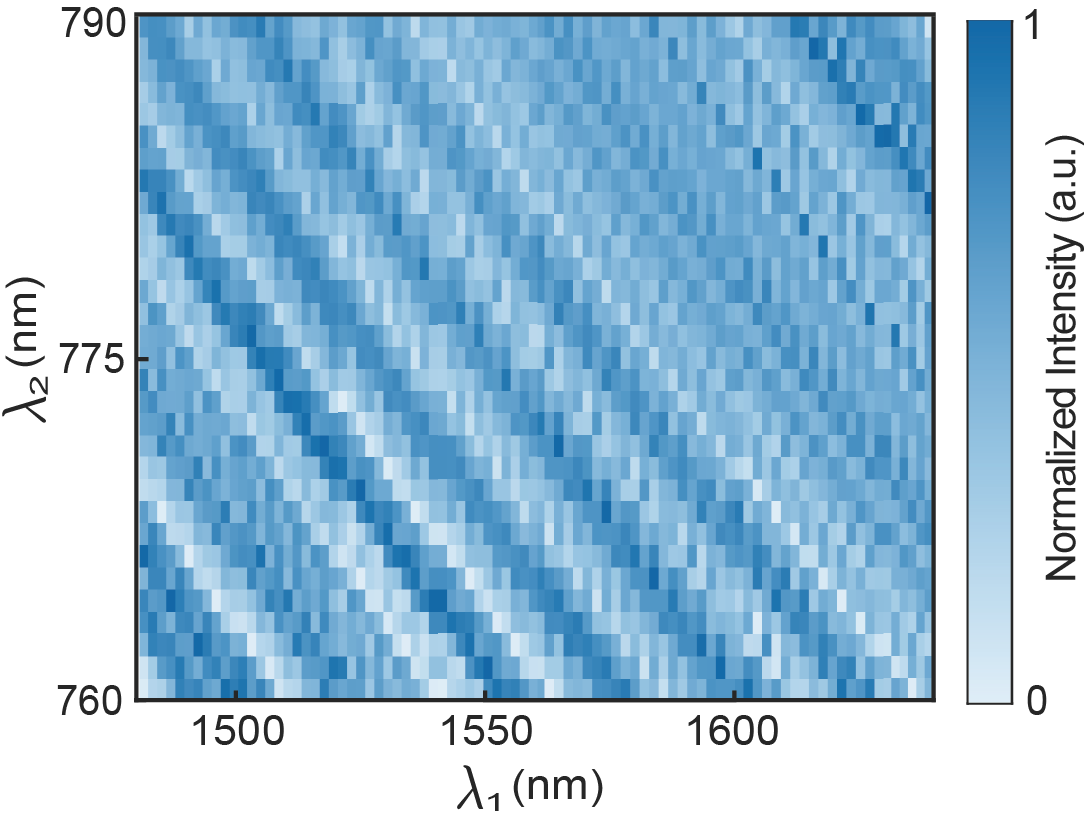}
\caption 
{Simulation of sum-frequency generation between the wavelengths around 780 nm and 1560 nm.}
\label{figs4}
\end{figure*} 
Based on the above theory, we simulate normalized SHG conversion efficiency of the device over a broad pump wavelength range (Fig.~\ref{figs3}). 
Despite a sharp dip near 1320~nm originating from a disruption of the phase-matching condition caused by a discontinuity of the segmented-chirped periodic poling, the device enables broadband and relatively efficient SHG.

We also simulate the SFG between the wavelengths around 780 nm and 1560 nm (Fig. \ref{figs4}), which shows broadband phase matching to generate light around third-harmonic wavelengths.
The anti-diagonal strips indicate the frequency correlation.

\section{Radio frequency beatnote}

\begin{figure*}[htbp] 
\includegraphics[width = 7in]{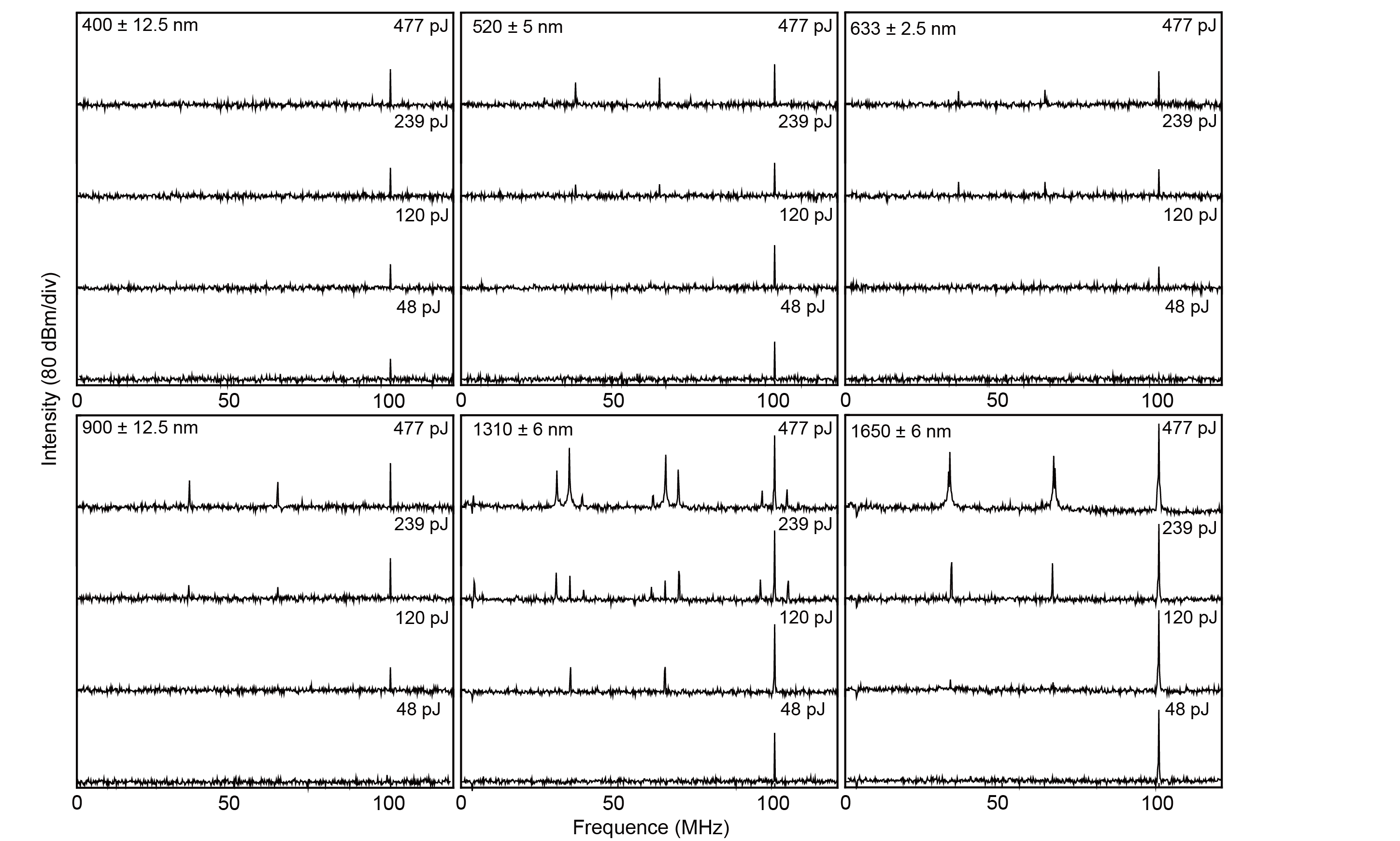}
\caption 
{RF beatnote measurements with increasing pump pulse energy (from bottom to top: 48 pJ, 120 pJ, 239 pJ and 477 pJ) at various wavelengths.}
\label{figs5}
\end{figure*} 

Lock-mode lasers generate an optical frequency comb that can be written as\cite{diddams2020optical}
\begin{equation}
f_m = m f_{\mathrm{rep}} + f_{\mathrm{ceo}},
\end{equation}
where $f_{\mathrm{rep}}$ is the pulse repetition rate, $f_{\mathrm{ceo}}$ is the carrier-envelope offset (CEO) frequency, and $m$ is the comb-line index. 
Due to the phase difference between the carrier wave and the envelope, the CEO frequency is usually nonzero. 
Starting from this fundamental comb, different nonlinear frequency-conversion processes can produce comb families with distinct CEO frequency characteristics.
For example, frequency comb generated at $n$-th harmonic wavelength can be generally expressed as\cite{fan2025spectral1}
\begin{equation}
f = m f_{\mathrm{rep}} + n f_{\mathrm{ceo}},
\end{equation}
where $n$ denotes the harmonic order, reflecting the order-dependent multiplication of the $f_{\mathrm{ceo}}$ in harmonic processes.

When comb families at different harmonics overlap spectrally, heterodyne beating between the corresponding comb lines gives rise to additional RF beat notes upon photodetection. 
The emergence of such RF beat notes indicates that at least two frequency-comb families with different offset characteristics coexist within the same spectral window, thereby providing direct evidence of multiple nonlinear interactions contributing to the overall frequency-conversion dynamics. 
Moreover, clear and stable RF beat notes are commonly regarded as a direct signature of good coherence in the generated supercontinuum \cite{wu2024visible1,fan2025spectral1}.

We record the RF beatnote spectra at various wavelengths of the supercontinuum, and the evolution of the RF beatnote signals with increasing pump energy can be clearly observed (Fig.~\ref{figs5}). 

\section{Comparison of 20 dB bandwidth of supercontinuum generation in various integrated photonic platforms.}
\begin{table*}[tbh!]
\begin{center}
\caption{\textbf{Comparison of on-chip supercontinuum generation} in terms of material platforms, device lengths ($L$), involved nonlinear interactions, pump wavelength ($\lambda_p$), pump pulse energy ($E$), and 20 dB bandwidth wavelength ranges of the output supercontinuum (BW$\rm _{20 dB}$).}
\label{tab}
\resizebox{0.95\textwidth}{!}{
\begin{tabular}{cccccc} 
\hline\hline
\rule[-1ex]{0pt}{3.5ex}  \textbf{Platform} & \textbf{\textit{L}} (mm) & \textbf{Interactions} & $\bm{\lambda_p}$ (\textmu m)& \textbf{\textit{E}} (pJ)& \textbf{BW$\bm{_{\rm20dB}}$} (nm)\\
\hline\hline
\rule[-1ex]{0pt}{3.5ex}  Si\cite{leo2014dispersive} & 7&$\chi^{(3)}$: SPM&1.565&4.8&~1420-1680\\
\hline
\rule[-1ex]{0pt}{3.5ex}  Si\cite{singh2018octave} & &$\chi^{(3)}$: SPM&1.95&16&~1600-2300\\
\hline
\rule[-1ex]{0pt}{3.5ex}  SiO$_2$\cite{yoon2017coherent} &15&$\chi^{(2)}$: SHG; $\chi^{(3)}$: SPM&1.064&2300&~900-1350\\
\hline
\rule[-1ex]{0pt}{3.5ex}  SiN$_x$\cite{grassani2019mid}&5 &$\chi^{(3)}$: SPM&2.09&530&~1500-2400, 3200-3800\\
\hline
\rule[-1ex]{0pt}{3.5ex}  SiN$_x$\cite{porcel2017two}&6 &$\chi^{(3)}$: SPM&1.56&1400&~1150-1850\\
\hline
\rule[-1ex]{0pt}{3.5ex}  SiC\cite{deniel2023visible} & 4.1&$\chi^{(2)}$: SHG; $\chi^{(3)}$: SPM&1.57&200&~ 1300-2100\\
\hline
\rule[-1ex]{0pt}{3.5ex}  AlN\cite{yan2026simplified}&6 & $\chi^{(3)}$: SPM &1.56&100&~1200-2300\\
\hline
\rule[-1ex]{0pt}{3.5ex}  AlGaAs\cite{kuyken2020octave}&3 &$\chi^{(3)}$: SPM &1.55&9.7&~1450-1800\\
\hline
\rule[-1ex]{0pt}{3.5ex}  GaN\cite{fan2024supercontinua}&5&$\chi^{(2)}$: SHG; $\chi^{(3)}$: SPM&1.56&527&~900-2400\\
\hline
\rule[-1ex]{0pt}{3.5ex}  InGaP\cite{dave2015dispersive}&1.5 &$\chi^{(3)}$: SPM&1.55&1.7&~1300-1850\\
\hline
\rule[-1ex]{0pt}{3.5ex}  Ta$_{2}$O$_{5}$\cite{jung2021tantala}&5&$\chi^{(2)}$: SHG; $\chi^{(3)}$: SPM&1.56&100&~1300-1800\\
\hline
\rule[-1ex]{0pt}{3.5ex}  TeO$_{2}$\cite{singh2020nonlinear}&7 &$\chi^{(3)}$: SPM, THG&1.55&84&~1420-1680\\
\hline
\rule[-1ex]{0pt}{3.5ex}  TiO$_{2}$\cite{hammani2018octave}&22&$\chi^{(3)}$: SPM&1.64&117&~1050-1910\\
\hline
\rule[-1ex]{0pt}{3.5ex}  As$_{2}$S$_{3}$\cite{hwang2021supercontinuum}&10 &$\chi^{(3)}$: SPM&1.56&77&~950-2900\\
\hline
\rule[-1ex]{0pt}{3.5ex}  GeSbS\cite{choi2016nonlinear}&15&$\chi^{(3)}$: SPM&1.55&170&~1350-2050\\
\hline
\rule[-1ex]{0pt}{3.5ex}  LT\cite{wang2025three}& 5 &$\chi^{(2)}$: SHG; $\chi^{(3)}$: SPM, THG &1.56&323&~1320-1850\\
\hline
\rule[-1ex]{0pt}{3.5ex} LN\cite{peng2025three}& 8&$\chi^{(2)}$: SHG; $\chi^{(3)}$: SPM, THG, HHG &1.4&$8*10^{5}$&~400-2500\\
\hline
\rule[-1ex]{0pt}{3.5ex}  LN\cite{hamrouni2024picojoule}&  &$\chi^{(2)}$: SHG,SFG; $\chi^{(3)}$: SPM &2.09&40&~700-2400\\
\hline
\rule[-1ex]{0pt}{3.5ex}  LN\cite{li20262}& 6.5 &$\chi^{(2)}$: SHG, SFG; $\chi^{(3)}$: SPM, THG &1.55&687&~1150-2050 \\
\hline
\rule[-1ex]{0pt}{3.5ex}  LN\cite{zhou2025quadratic}& 7 &$\chi^{(2)}$: SHG, DFG, OPA; $\chi^{(3)}$: SPM &2.09&177&~900-1400, 1600-2550\\
\hline
\rule[-1ex]{0pt}{3.5ex} LN\cite{jankowski2020ultrabroadband}& 6&$\chi^{(2)}$: SHG; $\chi^{(3)}$: SPM, THG, HHG &2.05&11.2&~500-750, 850-1300, 1550-2150\\
\hline
\rule[-1ex]{0pt}{3.5ex}  LN\cite{yu2019coherent}& 5 &$\chi^{(2)}$: SHG, SFG; $\chi^{(3)}$: SPM, THG &1.506&185&~1200-1700\\
\hline
\rule[-1ex]{0pt}{3.5ex}  LN\cite{fan2025spectral}& 4.5&$\chi^{(2)}$: SHG, SFG, DFG; $\chi^{(3)}$: SPM, THG&1.55&347&~1400-1800 \\
\hline
\rule[-1ex]{0pt}{3.5ex}  LN\cite{lu2019octave}& 10&$\chi^{(2)}$: SHG, SFG, DFG; $\chi^{(3)}$: SPM, THG&1.56&800&~1300-2100 \\
\hline
\rule[-1ex]{0pt}{3.5ex}  LN\cite{wu2024visible}& 6.6&$\chi^{(2)}$: SHG, SFG; $\chi^{(3)}$: SPM &1.55&90&~380-950, 1350-1900\\
\hline
\rule[-1ex]{0pt}{3.5ex}  LN\cite{Tang:25}& 20&$\chi^{(2)}$: SHG, SFG; $\chi^{(3)}$: SPM, THG &1.55&628&~900-2500\\
\hline
\rule[-1ex]{0pt}{3.5ex}  LN\cite{10.1063/5.0028776}&14&$\chi^{(2)}$: SHG; $\chi^{(3)}$: SPM&0.95&67&~850-1100\\
\hline
\rule[-1ex]{0pt}{3.5ex}  LN\cite{ayhan2025fabrication}& 4.5&$\chi^{(2)}$: SHG, SFG, DFG; $\chi^{(3)}$: SPM, THG&1.55&300&~1400-1700\\
\hline
\rule[-1ex]{0pt}{3.5ex}  LN\cite{ludwig2025mid}& 5&$\chi^{(2)}$: SHG, DFG ; $\chi^{(3)}$: SPM&1.55&236&~500-1100, 1500-2200\\
\hline
\rule[-1ex]{0pt}{3.5ex}  LN\cite{tang2025chip}& 6&$\chi^{(2)}$: SHG; $\chi^{(3)}$: THG, SPM &1.56&195&~1300-2000\\
\hline
\rule[-1ex]{0pt}{3.5ex} LN\cite{gao2025tightly}& 300&$\chi^{(2)}$: SHG; $\chi^{(3)}$: SPM &1.56&207&~1500-2000\\
\hline
\rule[-1ex]{0pt}{3.5ex}  This work& 14&$\chi^{(2)}$: SHG, SFG, OPA; $\chi^{(3)}$: SPM, THG, HHG&1.56&477&320-2600\\
\hline\hline
\end{tabular}
}
\end{center}
\end{table*}

\newpage 
\bibliographystyle{ieeetr}  

\bibliography{SI_ref}